\documentclass[fleqn,usenatbib]{mnras}

\usepackage{newtxtext,newtxmath}

\usepackage[T1]{fontenc}

\DeclareRobustCommand{\VAN}[3]{#2}
\let\VANthebibliography\thebibliography
\def\thebibliography{\DeclareRobustCommand{\VAN}[3]{##3}\VANthebibliography}

\usepackage{graphicx}
\usepackage{amsmath}
\usepackage{caption}

\title[The galaxy UV LF at $z = 8-15$]{The evolution of the galaxy UV luminosity function at redshifts $\mathbf {z \simeq 8}$~--~$\mathbf{15}$ from deep JWST and ground-based near-infrared imaging}

\author[C.\,T. Donnan et al.]{C.\,T. Donnan$^{1}$\thanks{E-mail: callum.donnan@ed.ac.uk},
D.\,J. McLeod$^{1}$,
J.\,S. Dunlop$^{1}$,
R.\,J. McLure$^{1}$,
A.\,C. Carnall$^{1}$,
R. Begley$^{1}$,
F. Cullen$^{1}$ 
\newauthor
M.\,L. Hamadouche$^{1}$, R.\,A.\,A. Bowler$^{2}$, D.\,Magee$^{3}$, H.\,J. McCracken$^{4}$, B. Milvang-Jensen$^{5,\,6}$ 
\newauthor
A. Moneti$^{4}$ \& T. Targett$^{7}$
\\
$^{1}$Institute for Astronomy, University of Edinburgh, Royal Observatory, Edinburgh, EH9 3HJ. UK\\
$^{2}$Jodrell Bank Centre for Astrophysics, University of Manchester, Oxford Road, Manchester, UK\\
$^{3}$ Department of Astronomy and Astrophysics, UCO/Lick Observatory, University of California, Santa Cruz, CA 95064, USA\\
$^{4}$ Institut d’Astrophysique de Paris, UMR 7095, CNRS, and Sorbonne 
Universit\'{e}, 98 bis boulevard Arago, 75014 Paris, France\\
$^{5}$ Cosmic Dawn Center (DAWN)\\
$^{6}$ Niels Bohr Institute, University of Copenhagen, Jagtvej 128, 2200 Copenhagen, Denmark\\
$^{7}$ Department of Physics and Astronomy, Sonoma State University, 1801 East Cotati Avenue, Rohnert Park, CA 94928-3609, US
}

\date{Accepted XXX. Received YYY; in original form ZZZ}

\pubyear{2022}

\begin{document}
\label{firstpage}
\pagerange{\pageref{firstpage}--\pageref{lastpage}}
\maketitle

\begin{abstract}
We reduce and analyse the available \textit{James Webb Space Telescope} (JWST) ERO and ERS NIRCam imaging (SMACS0723, GLASS, CEERS) in combination with the latest deep ground-based near-infrared imaging in the COSMOS field (provided by UltraVISTA DR5) to produce a new measurement of the evolving galaxy UV luminosity function (LF) over the redshift range $z = 8 - 15$. This yields a new estimate of the evolution of UV luminosity density ($\rho_{\rm UV}$), and hence cosmic star-formation rate density ($\rho_{\rm SFR}$) out to within $< 300$\, Myr of the Big Bang. Our results confirm that the high-redshift LF is best described by a double power-law (rather than a Schechter) function up to $z\sim10$, and that the LF and the resulting derived $\rho_{\rm UV}$ (and thus $\rho_{\rm SFR}$), continues to decline gradually and steadily up to $z\sim15$ (as anticipated from previous studies which analysed the pre-existing data in a consistent manner to this study). We provide details of the 61 high-redshift galaxy candidates, 47 of which are new, that have enabled this new analysis. Our sample contains 6 galaxies at $z \ge 12$, one of which appears to set a new redshift record as an apparently robust galaxy candidate at $z \simeq 16.4$, the properties of which we therefore consider in detail. The advances presented here emphasize the importance of achieving high dynamic range in studies of early galaxy evolution, and re-affirm the enormous potential of forthcoming larger JWST programmes to transform our understanding of the young Universe.  
\end{abstract}

\begin{keywords}
galaxies: evolution -- galaxies: formation -- galaxies: high redshift
\end{keywords}

\section{Introduction}

Over the last decade, major instrumental advances have enabled astronomers to clarify the background cosmology of the Universe 
and push studies of galaxies back to within a billion years 
of the Big Bang (see \citet{dunlop2013}, \citet{madau2014} \&
\citet{stark2016} for reviews). In particular, deep near-infrared extragalactic surveys, both from the ground and with the \textit{Hubble Space Telescope} (HST) and \textit{Spitzer}
have revealed galaxy evolution extending out to redshifts $z \simeq 10$
\citep[e.g.][]{ellis2013,mclure2013,oesch2014,oesch2018,bowler2014,bowler2015,bowler2020,finkelstein2015,mcleod2015,mcleod2016,bouwens2021,bouwens2022}.
It now appears that this growing population of early star-forming galaxies could indeed have bathed the Universe in sufficient high-energy photons to produce cosmic hydrogen reionization \citep{robertson2015,finkelstein2019,aird2015}, especially since $\mu$-wave background measurements now indicate a `mean' redshift of reionization $\langle z \rangle \simeq 7.8 \pm 0.7$ \citep{planck2020}.
However, analyses combining all available constraints indicate that
much of the key `action' has yet to be discovered, with the ionizing photon
budget potentially dominated by low-luminosity galaxies undetected by HST, and early galaxies commencing 
reionization at redshifts as high as $z > 15$ \citep{robertson2015,robertson2021}. Testing these predictions is one key goal for the
\textit{James Webb Space Telescope} (JWST), given its ability to probe out to $z \simeq 20$.

Although installation of the near-infrared camera WFC3/IR enabled HST to probe beyond $z \simeq 7$ into the first $\simeq$ Gyr, the isolation of secure samples of $z > 7$ galaxies has still been severely hampered by the curtailment of
HST wavelength coverage at $\lambda_{\rm obs} < 1.6$\,$\mu$m. Robust redshift information benefits greatly not only from identification
of the Lyman-break at $\lambda_{\rm rest} = 1216$\,\AA, but also from extended/high-quality wavelength coverage above the 
break to exclude lower-redshift red/dusty interlopers or extreme emission-line objects which can masquerade as very high redshift Lyman-break galaxies. 
This uncertainty beyond $z \simeq 8$ explains, at least in part, why different authors have reached very different conclusions regarding the very high-redshift evolution of the galaxy LF, and hence the high-redshift decline of cosmic star-formation rate density ($\rho_{\rm SFR}$). In particular, based on deep HST surveys, \citet{mcleod2015,mcleod2016} concluded in favour of a smooth, exponential decline in $\rho_{\rm SFR}$ out to at least $z \simeq 10$, whereas \citet{oesch2013,oesch2014,oesch2018} deduced the existence of a much more rapid decline/cutoff beyond $z \simeq 8$; the implications of these alternative forms of high-redshift evolution for galaxy formation, and for the prospects of finding galaxies at extreme redshifts, are very different. 

For probing beyond $z \simeq 7$, and resolving such uncertainties, the capabilities of the NIRCam camera on-board JWST are transformative, with complete multi-band imaging now available out to $\lambda \simeq 5$\,$\mu$m with unprecedented angular resolution. 

Studies of high-redshift galaxy evolution with HST have also been limited by the areal coverage of near-infrared HST imaging, due to the small field-of-view of the WFC3 camera. Consequently, even though heroic efforts have been made to construct large-area HST image mosaics \citep[e.g., CANDELS:][]{grogin2011}, degree-scale near-infrared imaging from the ground, in particular with the WFCAM camera on UKIRT \citep{lawrence2007}, and VIRCAM on VISTA \citep{mccracken2012}, has continued to drive our knowledge of the evolution of the brightest galaxies at $z > 6$ \citep{bowler2014,bowler2015}. Indeed, the dynamic range (in galaxy luminosity) which can be accessed by combining space-based and ground-based near-infrared galaxy surveys has proved to be invaluable/essential for constraining the evolving form of the galaxy UV luminosity function out to the highest redshifts, and hence enabling meaningful comparison with theoretical models of early galaxy evolution \citep{bowler2020,adams2022}. 

It is the power of this combined ground-based + space-based approach which we exploit again in this new study of the high-redshift galaxy LF, but now for the first time armed with $\simeq 45$\,arcmin$^2$ of deep multi-band NIRCam imaging from the JWST Early Release Observations (ERO) and Early Release Science (ERS) programmes, and $\simeq 1.8$\,deg$^2$ of near-homogeneous near-infrared imaging in the COSMOS field provided by Data Release 5 (DR5) from the UltraVISTA survey \citep{mccracken2012}. We have used these brand-new datasets to conduct a new search for galaxies at $z\geq7.5$ in the three early JWST deep fields (SMACS0723, CEERS and GLASS) and in the COSMOS/UltraVISTA field. The resulting new high-redshift galaxy samples have enabled us to derive a new estimate of the evolution of the galaxy LF, and hence $\rho_{\rm SFR}$, from $z \simeq 8$ out to $z \simeq 15$, less that 300\,Myr after the Big Bang.

The paper is structured as follows. In Section \ref{sec:data} we describe the JWST and ground-based data which we utilised in each field. In Section \ref{sec:catalogue} we describe the process of catalogue creation and galaxy selection which yielded the new galaxy sample presented in Section \ref{sec:sample}. Then, in Section \ref{sec:LF} we describe how the luminosity function was calculated and present our new determination of the evolving, 
high-redshift, galaxy UV LF, along with the resulting constraints on UV luminosity density, $\rho_{\rm UV}$, and hence cosmic star-formation rate density, $\rho_{\rm SFR}$ out to $z \simeq 15$. Finally, in Section \ref{sec:discussion} we discuss our results in the context of existing studies, before summarising our conclusions in Section \ref{sec:conclusions}. Throughout we use magnitudes in the AB system \citep{oke1974,oke1983}, and assume a standard cosmological model with $H_0=70$ km s$^{-1}$ Mpc$^{-1}$, $\Omega_m=0.3$ and $\Omega_{\Lambda}=0.7$.

\section{Data}
\label{sec:data}

\subsection{Fields}

\subsubsection{JWST Early Release imaging}
We utilise the early deep public imaging data from JWST covering three separate fields. First, the SMACS J0723 cluster was imaged using NIRCam in the F090W, F150W, F200W, F277W, F356W, F444W filters as part of the Early Release Observations (ERO) \citep{pontoppidan2022}. This imaging dataset consists of a single NIRCam pointing which targets the cluster with one NIRCam module while the other module delivers imaging in a blank field adjacent to the cluster (hence yielding a "parallel", relatively unlensed survey field). Second, the CEERS Early Release Science (ERS) programme has now observed 4 of the 10 planned NIRCam pointings in the Extended Groth Strip (EGS) CANDELS field, and here we use the resulting NIRCam imaging in the F115W, F150W, F200W, F277W, F356W, F410M and F444W filters. Finally, the GLASS ERS programme \citep{treu2022} has already yielded a parallel NIRCam field consisting of one (two module) pointing imaged in the F090W, F115W, F150W, F200W, F277W, F356W, and F444W filters. However, the GLASS F090W imaging contains an abundance of artefacts across the image making it challenging for use in searching for $z\geq7.5$ galaxies, and so in this study, for high-redshift galaxy selection, we utilise only the F115W, F150W, F200W, F277W, F356W and F444W imaging (although see \citet{leethochawalit2022b}). The final combined JWST NIRCam effective imaging area available for this study totals $\simeq 45$\,arcmin$^2$, albeit with the somewhat varied filter coverage described above. This public JWST NIRCam imaging was reduced using \texttt{PENCIL} (PRIMER enhanced NIRCam Image Processing Library). The \texttt{PENCIL} pipeline is built on top of STScI's JWST Calibration (v1.6.2) but also includes additional processing steps not included in the standard calibration pipeline. This includes the subtraction of 1/f noise striping patterns (both vertical and horizontal) that are not fully removed by the standard calibration pipeline and the subtraction of ``wisps'' artifacts from the short wavelength filters F150W and F200W in the NRCA3, NRCB3, and NRCB4 detector images.

Additionally, the background sky subtraction is performed by subtracting the median background over a NxN grid while using a segmentation map to mask pixels attributed to sources. The image alignment is executed in two passes using the calibration pipeline’s TweakReg step and then using STScI python package TweakWCS: the first pass uses TweakReg to group overlapping images for each detector/filter and perform an internal alignment within the detector/filter group; the second performs alignment against an external catalog using TweakWCS. The external catalog is, if possible, generated from an HST ACSWFC image mosaic which has been registered to the GAIA DR3 catalog.
The astrometry of all the reduced images was aligned using \texttt{SCAMP} to GAIA EDR3 and aligned and stacked to the same pixel scale of 0.03\,arcsec using \texttt{SWARP}.

\subsubsection{COSMOS/UltraVISTA}
We utilise near-infrared imaging from the UltraVISTA survey \citep{mccracken2012} which provides deep $YJHK_s$ imaging across 1.8\,deg$^2$ in the COSMOS field, taken using ESO's VISTA telescope in Chile. The UltraVISTA imaging is split into two regions ``ultra-deep" and ``deep" which cover approximately half the area each. These regions consist of four stripes, each of which alternate between the two depths across the image. In this study we use the fifth data release (DR5) of UltraVISTA which differs primarily from DR4 in providing significant deeper $J$-band and $H$-band imaging: $\simeq 1$\,mag. deeper in the ``deep" stripes, and $\simeq 0.2$ mag. deeper in the ``ultra-deep" stripes. Within the central 1\,deg$^2$ we complement the new near-infrared UltraVISTA imaging data with optical imaging from the CFHTLS-D2 field from the CFHT Legacy Survey \citep{hudelot2012} in $u^*griz$. We also include wider-area optical data covering the full UltraVISTA field from the Hyper Suprime-Cam Subaru Strategic Program (HSC-SSP) DR2 \citep{aihara2019} in the $GRIZy$ filters as well as in two narrow bands, $NB816$ and $NB921$. All the near-infrared and optical imaging in COSMOS was aligned to the GAIA EDR3 reference frame using \texttt{SCAMP} and re-sampled using \texttt{SWARP} to a common pixel scale of 0.15-arcsec. Finally, we supplemented our ground-based datasets by adding  $3.6\mu$m and $4.5\mu$m photometry from \textit{Spitzer}/IRAC imaging, which experience proves can be invaluable for the refinement of photometric redshifts as well as minimising the level of low-redshift galaxy and dwarf-star contamination in the final high-redshift galaxy sample. The \textit{Spitzer}/IRAC data in 3.6 $\mu$m and 4.5 $\mu$m in the COSMOS field was provided by the Cosmic Dawn Survey \citep{moneti2022}.

\subsection{Image processing}

\subsubsection{PSF homogenisation in COSMOS}
In order to derive consistent photometry in different filters, the differences in the point spread function (PSF) between filters needs to be accounted for. We corrected for this in the COSMOS imaging by homogenising the PSFs in the different images to one common PSF. As the UltraVISTA $Y$ band has the broadest PSF, we chose to use that as the target PSF to which to PSF-homogenise all of the other COSMOS imaging. Firstly, we identified $\simeq 15$ bright but unsaturated stars in each image. We then centroided and stacked these stars to generate a measurement of the PSF in each waveband. Then, using a combination of a Moffat profile with two Gaussian profiles, we generated a series of kernels. These kernels were then convolved with the original PSFs to match the target ($Y$-band) PSF. At a radius of 0.9-arcsec we confirmed that the enclosed flux in every image is within $2\%$ of the target. We then convolved every image with its respective kernel to PSF-homogenise the entire COSMOS imaging dataset.

\subsubsection{Image depths}
The global depths in all of the ground-based PSF homogenised images were determined using 1.8-arcsec diameter circular apertures placed in all locations within the image that were determined to be source free. The $5\sigma$ depth was then calculated via
\begin{equation}
    5\sigma = 1.483 \times \rm{MAD} \times 5,
    \label{eq:depth}
\end{equation}
where MAD refers to the median absolute deviation of the flux detected in the empty apertures.
These \textit{global} $5$-$\sigma$ depths for each ground-based image are listed in Table~\ref{tab:depths} for information. In practice we then re-determined \textit{local} depths for each source detected (see Section~\ref{sec:catalogue}) by determining equation~(\ref{eq:depth}) on the 200 empty apertures nearest to the source in question, and adopted the $1\sigma$ local depth as the uncertainty in the photometry for every source detected.

\begin{table}
	\centering
	\caption{The derived $5\sigma$ global depths for all the COSMOS images used in this analysis. All depths (quoted in AB Magnitudes) were calculated using 1.8-arcsec diameter apertures on the PSF-homogenised images and corrected to total using a point-source correction.}
	\label{tab:depths}
	\begin{tabular}{lcr} 
		\hline
		Filter & ultra-deep & deep\\
		\hline
        CFHT $u^*$  &     27.00 &     27.00 \\
        CFHT $g$ &  27.03 &  27.03 \\
        CFHT $r$  &  26.47 &  26.47 \\
        CFHT $i$  &  26.17 &  26.17 \\
        CFHT $z$ &  25.34 &  25.34 \\
        SSC $B$ &  27.16 &  27.16 \\
        SSC $z \prime_{new}$  &  25.95 &  25.95 \\
        HSC $G$ &  27.09 &  27.09 \\
        HSC $R$ &  26.74 &  26.74 \\
        HSC $I$ &  26.46 &  26.46 \\
        HSC $Z$  &  26.18 &  26.18 \\
        HSC $y$  &  25.42 &  25.42 \\
        HSC $NB816$  &  25.66 &  25.66 \\
        HSC $NB921$  &   25.70 &   25.70 \\
        VISTA $Y$  &  25.51 &  24.37 \\
        VISTA $J$ &  25.55 &   25.10 \\
        VISTA $H$  &  25.26 &  24.96 \\
        VISTA $K_s$  &  24.96 &  24.62 \\
		\hline
	\end{tabular}
\end{table}

For the JWST images, the global depths of the SW images were determined using 0.248-arcsec diameter circular apertures placed in all locations within the image that were determined to be source free. We used the same procedure for the LW images, but with 0.341-arcsec diameter apertures. The $5$-$\sigma$ depth was then calculated using the procedure described above (equation~(\ref{eq:depth})) . The  \textit{global} $5$-$\sigma$ depths (corrected to total with the appropriate point-source correction) for all three JWST fields are shown in Table \ref{tab:depths_jwst}. In practice, we again determined local depths for each source detected using the 200 empty photometric apertures closest to the source. We adopted the (point-source corrected) $1$-$\sigma$ local depth as the uncertainty on our photometry for every source detected. 

\begin{table}
	\caption{The derived $5\sigma$ global depths for all the space-based images used in this analysis. All depths (given in AB magnitudes) have been corrected to total assuming a point-source correction.}
	\label{tab:depths_jwst}
	\begin{tabular}{lcccr} 
		\hline
	     & SMACS & SMACS &  & \\
		Filter & Cluster & Parallel & CEERS & GLASS\\
		\hline
		F090W & 28.30 & 28.40 & - & -\\
		F115W & - & - & 28.62 & 28.75\\
		F150W & 28.37 & 28.70 & 28.54 & 28.57\\
		F200W & 28.33 & 28.78 & 28.70 & 28.67\\
		F277W & 27.65 & 28.86 & 28.74 & 28.77\\
		F356W & 28.05 & 28.89 & 28.77 & 28.75\\
		F410M & - & - & 29.07 & -\\
		F444W & 28.27 & 28.68 & 28.34 & 28.79\\

		\hline
	\end{tabular}
\end{table}

\subsubsection{\textit{Spitzer}/IRAC fluxes}
The \textit{Spitzer}/IRAC imaging at 3.6\,$\mu$m and 4.5\,$\mu$m has significantly poorer angular resolution than the optical and near-infrared imaging used in this study. Therefore, to extract robust IRAC photometry for the COSMOS field, we utilised the deconfusion software package \texttt{TPHOT} \citep{merlin2015}. We used the three near-infrared detection images (see below) as the high-resolution priors to generate the \texttt{TPHOT} fluxes which are therefore isophotal. To add this to the PSF-homogenized photometry, we performed a correction to the optical and near-infrared photometry by multiplying the fluxes by $f=D_{iso}/D_{1.8}$ where $D_{iso}$ is the isophotal flux in the corresponding detection image and $D_{1.8}$ is the flux enclosed within the 1.8-arcsec diameter aperture in the detection image. 

\section{Catalogue creation and galaxy selection}
\label{sec:catalogue}
To create catalogues in the COSMOS fields we utilised inverse variance weighted stacks of the data in the $Y$, $J$, $H$, and $K_s$ bands to increase the sensitivity of our detections. We constructed stacks of VISTA $Y + J + H + K_s$, VISTA $J + H + K_s$ and VISTA $H + K_s$ imaging. These stacks were chosen to best optimise the detection of $z\sim 6-10$ Lyman-break galaxies. The catalogues were created using SExtractor \citep{bertin1996} in dual-image mode with the stacked images as the detection images. A master catalogue was created by combining sources detected in both the detection images, with duplicates removed by retaining the object with the highest signal-to-noise. 

For the JWST catalogues we created two rest-frame UV-selected catalogues using SExtractor in dual-image mode with the F200W as the detection image. This was to optimise our catalogue to select $z\geq8$ galaxies as this filter will encompass the bright UV flux red-ward of the Lyman break. We used 8-pixel diameter (0.248-arcsec diameter) apertures on the imaging taken through the SW filters (F090W, F115W, F150W, F200W) and 11-pixel diameter (0.341-arcsec diameter) apertures on the LW imaging (F277W, F356W, F410M, F444W) as these diameters were found to contain a similar percentage of total flux based on the curve-of-growth ($\sim$76\%). Further small corrections on the percent level were made to correct to 76\% of total flux based on a point-source correction derived from curves of growth determined from the imaging in each JWST NIRCam filter.

\subsection{Determination of photometric redshifts}
We used the photometric redshift (photo-z) code \texttt{EAZY} \citep{brammer2008} for our redshift determination for every object in the COSMOS and JWST catalogues. We ran \texttt{EAZY} using the Pegase set of templates with zero-point offsets calculated based on a set of robust spectroscopic redshifts in COSMOS. This method allows us to refine the SED fitting using spectroscopically confirmed redshifts and assess the performance of the SED fitting by calculating the fraction of catastrophic outliers ($f_{\rm outliers}$) and the bias, which we define as the median value of $\rm{d}z = (z_{\rm spec} - z_{\rm phot})/(1+z_{\rm spec})$. 
To quantify the accuracy of the photometric redshifts we calculated $\sigma_{\rm{d}z}$ using the robust median absolute deviation (MAD) estimator.
A comparison of the spectroscopic redshifts vs photometric redshifts for $\sim3700$ sources shows that our photometric redshifts are robust, with a $\sigma_{\rm{d}z}=0.025$ and an outlier rate of $f_{\rm outliers}=2.49\%$. We did not initially include the \textit{Spitzer}/IRAC photometry in the fitting, as this was added after the initial galaxy selection described in Section \ref{sec:selection} to refine the selected sample.  Because brown dwarfs are possible contaminants in the search for high-redshift galaxies (especially with ground-based data), we also used \texttt{EAZY} to fit a series M-, L- and T-dwarf templates from the SpeX prism library\footnote{http://pono.ucsd.edu/$\sim$adam/browndwarfs/spexprism/index.html} to the COSMOS catalogues. 

\subsection{Galaxy selection from the COSMOS UltraVISTA imaging}
\label{sec:selection}
From the sample of objects detected in the COSMOS field we selected galaxies in redshift bins of width $\Delta z=1$ around central redshifts of $z = 8, 9$ and $10$. For a source to be accepted into the sample, they must meet the following criteria:

\begin{enumerate}
    \setlength\itemsep{1em}
    \item $\chi^2_{\nu,\rm{galaxy}} < 5$
    \item $\chi^2_{\nu,\rm{galaxy}}<\chi^2_{\nu,\rm{star}}$
\end{enumerate}

\noindent
where $\chi^2_{\nu, \rm{galaxy}}$ and $\chi^2_{\nu, \rm{star}}$ represent the reduced $\chi^2$ for the Pegase galaxy templates, and SpeX stellar templates, respectively. Condition (i) ensures that only sources with acceptable galaxy template solutions are included. Condition (ii) removes brown dwarf contaminants by ensuring that the galaxy templates provide a better fit than the brown dwarf templates. 
 
To further refine the sample we use two further SED fitting codes: \texttt{LePhare} \citep[][]{arnouts1999,ilbert2006} with templates from \citet{bruzual2003} and with dust attenuation spanning the range $A_{V}=0.0-6.0$, and the code described in \citet{mclure2011}. We further require all galaxies to have a preferred high-redshift solution produced by these two alternative codes, to ensure that the redshift solution is robust against choice of templates, dust attenuation and photo-z code.

Finally, all candidates were visually inspected to remove objects which could be due to diffraction spikes and any other artefacts.

\subsubsection{Cross-talk artefacts}
In the COSMOS field, \citet{bowler2017} identified faint cross-talk in the VISTA $YJHK_s$ imaging. Therefore, to avoid these artefacts we developed a mask based on the positions of all the bright stars in the image from the COSMOS 2020 bright stars mask \citep{weaver2022}. 

\subsubsection{z=8}
In COSMOS we require a $5\sigma$ detection in the VISTA $J$ or $H$ band. We require non-detections at the $2\sigma$ level in all filters blue-ward of the Lyman break up to and including the SSC $z^\prime$ filter. A best-fitting photo-z in the range $7.5<z<8.5$ from \texttt{EAZY} is also required. 

\subsubsection{z=9}
In COSMOS we require a $5\sigma$ detection in the VISTA $J$, $H$ or $K_s$ band. We require non-detections at the $2\sigma$ level in all filters blue-ward of the Lyman break up to and including the VISTA $Y$ filter. A best-fitting photo-z in the range $8.5<z<9.5$ from \texttt{EAZY} is also required. 

\subsubsection{z=10}
In COSMOS we require a $5\sigma$ detection in the VISTA $H$ or $K_s$ band. We require non-detections at the $2\sigma$ level in all filters blue-ward of the Lyman break up to and including the VISTA $Y$ filter. A best-fitting photo-z in the range $9.5<z<10.5$ from \texttt{EAZY} is also required. 

\subsection{Galaxy selection from the JWST NIRCam imaging}
We selected galaxies using different `dropout' criteria in the JWST fields. Due to the different filter sets in the three different fields, the same criteria could not be applied to every field. The conditions required to select robust samples of high-redshift galaxies are therefore described below, field by field.

\subsubsection{CEERS and GLASS} 
In the CEERS and GLASS fields we constructed three samples meeting the following criteria.
F115W dropouts require a $2\sigma$ non-detection in F115W with a $5\sigma$ detection in F150W and a $3\sigma$ detection in F200W. F150W dropouts were selected by requiring a $2\sigma$ non-detection in F115W and F150W, a $5\sigma$ detection in F200W and a $3\sigma$ detection in F277W. We also included sources where the Lyman break is partway through the F150W filter: this sample requires a $2\sigma$ non-detection in F090W and F115W, a detection between $2\sigma$ and $5\sigma$ in F150W, a $5\sigma$ detection in F200W and a $3\sigma$ detection in F277W.

\subsubsection{SMACS0723}
We performed different dropout selections for sources in SMACS0723 due to the inclusion of F090W imaging and the lack of F115W imaging. F090W dropouts require a $2\sigma$ non-detection in F090W with a $5\sigma$ detection in F150W and a $3\sigma$ detection in F200W. F150W dropouts require $2\sigma$ non-detections in F090W and F150W with a $5\sigma$ detection in F200W and a $3\sigma$ detection in F277W.

For SMACS0723, CEERS and GLASS, every galaxy in the final selected sample was also required to have a $\Delta \chi^2>4$ between the best fitting high-$z$ and low-$z$ solution. This helps to ensure that the  high-$z$ solution is robust by removing potential low-$z$ contaminants (generally dusty intermediate-redshift galaxies or extreme emission-line objects) from the sample. Consistent with the selection of the ground-based galaxies, we also fitted the sources in the JWST sample with the SED code \texttt{LePhare} as well as the code described in \citet{mclure2011}, and additionally required preferred high-redshift solutions from both codes. Finally, all candidates were again visually inspected to remove artefacts.

\section{The final high-redshift galaxy sample}
\label{sec:sample}

\subsection{The final COSMOS/UltraVISTA galaxy sample}

Using the selection criteria described in Section \ref{sec:selection} we assembled a final combined sample of 16 LBGs at $z>7.5$ in COSMOS/UltraVISTA field. The objects with their best fitting redshift, $M_{\rm UV}$ and coordinates are listed in Table \ref{tab:sample} (ranked by photometric redshift). The sample contains 12 sources with a best fitting redshift from \texttt{EAZY} in the range $7.5<z<8.5$, 3 sources in the range $8.5<z<9.5$ and 1 source in the range $9.5<z<10.5$. 

\begin{table*}
	\centering
	\caption{The best-fitting photometric redshifts from \texttt{EAZY} for the final sample of $z>7.5$ galaxies found in the COSMOS/UltraVISTA field, ranked by photometric redshift. The first column gives the source ID, with $z_{\rm phot}$ for each object then presented in Column 2. Column 3 gives the derived rest-frame UV magnitude of each galaxy. Column 4 denotes the the sub-region of the UltraVISTA imaging within which each object has been found: `U-D' refers to the ultra-deep stripes while  `D' refers to the deep stripes, although as discussed in the text the difference in depth between these two regions has now been largely eliminated at $J$,$H$,$K_s$ in UltraVISTA DR5. The coordinates for each source are given  in the following two columns. The final two columns list an alternative ID, as appropriate, for those (8) sources which were already detected by \citet{bowler2020} or were listed in the COSMOS2020 catalogue \citep{weaver2022}, indicated by B20 and W22, respectively. Interestingly, and as largely anticipated, the new sources reported here almost all lie with the D (deep) stripes, where the UltraVISTA data have been most improved in depth between DR4 and DR5 (effectively doubling the area available in COSMOS for the selection of very high-redshift galaxies).}
	\label{tab:sample}
	\renewcommand{\arraystretch}{1.35} 
	\begin{tabular}{lccccccr} 
		\hline
		ID & $z_{\rm phot}$ & $M_{\rm UV}$ & Region & RA & DEC & B20 & W22\\
		\hline
  334330 &    7.58$^{+0.87}_{-0.16}$ & $-21.30$   & D & 10:00:05.27 & 01:59:05.98 & - & -\\
  733875 &    7.58$^{+0.46}_{-0.31}$ & $-21.57$ & D & 09:59:52.85 & 02:34:57.00 & - & -\\
  812867 &    7.58$^{+1.35}_{-0.02}$ & $-21.02$  & U-D & 10:00:040.8 & 02:42:16.62 & - & 1349252\\
  688541 &    7.66$^{+0.82}_{-0.00}$ & $-22.15$   & U-D & 10:02:12.55 & 02:30:45.81 & 914 & 1151531\\
  765906 &    7.66$^{+0.53}_{-0.1}$ & $-22.61$   & D & 09:58:12.23 & 02:37:52.61 & - & 1274544 \\
  626972 &    7.75$^{+0.72}_{-0.24}$ & $-21.49$   & U-D & 09:57:54.25 & 02:25:08.40 & 839 & 1055131 \\
  536767 &    8.02$^{+0.33}_{-0.40}$ & $-21.40$   & D & 09:58:17.19 & 02:17:06.39 & - & -\\
  861605 &    8.02$^{+0.66}_{-0.26}$ & $-21.33$   & U-D & 09:57:21.37 & 02:45:57.57 & - & 1412106\\
  978389 &    8.02$^{+0.77}_{-0.21}$ & $-21.68$   & U-D & 10:00:34.56 & 01:55:17.42 & - & -\\
  484075 &    8.11$^{+1.03}_{-1.03}$ & $-22.05$   & D & 09:58:032.1 & 02:12:21.83 & - & -\\
  578163 &    8.20$^{+0.50}_{-0.35}$ & $-22.35$   & U-D & 09:57:47.91 & 02:20:43.54 & 762 & 978062 \\
  458445 &    8.38$^{+0.28}_{-0.56}$ & $-21.65$   & U-D & 10:01:47.49 & 02:10:15.43 & 598 & 784810\\
  448864 &    8.57$^{+0.30}_{-0.69}$ & $-21.15$   & D & 10:02:46.29 & 02:09:23.42 & - & -\\
  306122 &    $8.76^{+0.14}_{-0.43}$ & $-21.76$   & D & 10:02:50.81 & 01:56:36.49 & - & -\\
  892014 &    8.96$^{+0.11}_{-0.33}$ & $-22.16$   & D & 10:00:04.23 & 02:47:59.84 & - & -\\
  817482 &    9.89$^{+1.22}_{-0.20}$ & $-22.57$   & U-D & 09:57:25.46 & 02:42:41.21 & - & 1356755\\

		\hline
	\end{tabular}
\end{table*}

There are 16 galaxies in this sample, 8 of which are found in the "Deep" region of UltraVISTA. In previous work searching for bright galaxies at $z>7.5$ in UltraVISTA, the fraction of the sample that was in the "Deep" region was minimal with $1/16$ in the \citet{bowler2020} sample, and $0/16$ in the \citep{stefanon2019} sample.

This demonstrates the (anticipated) impact of the increased depth in the $J$ and $H$ bands delivered in the "Deep" region in UltraVISTA DR5, which effectively doubles the useful area searchable for bright high-redshift galaxies in the COSMOS field. 

\subsection{The final JWST galaxy sample}
\label{sec:jwst_sample}
There are 45 galaxies in the final high-redshift ($z > 8.5$) sample uncovered by our analysis of the ERO/ERS JWST NIRCam imaging, with 23 found in SMACS0723, 19 in CEERS, and only 3 in GLASS. Two of the high-redshift galaxies we have found here in the GLASS field have been independently discovered by \citet{naidu2022} and \citet{castellano2022}, namely GLASS-1698 and GLASS-17487 in our sample which we find to lie at $z=10.45$ and $z=12.42$ respectively (consistent with the independently reported photometric redshifts). We fail to recover robust high-redshift solutions for any of the five sources that \citet{castellano2022} reported in their fainter sample. 
In SMACS0723, 2 of the $z_{\rm{phot}}\sim9$ galaxies in this sample are known to be at $z<8.5$ as they have been spectroscopically confirmed at $z=7.663$ (SMACS-44711) and $z=7.665$ (SMACS-44566) as noted in \citet{carnall2022}.
Therefore, these sources are not included in the LF calculation described in Section \ref{sec:number_density}.
We have also detected the $z\sim12$ source discovered in \citet{finkelstein2022} with a similar redshift of $z=12.29$ (ID: CEERS-32395\_2).
The highest redshift galaxy in our sample is CEERS-93316 which sets a new redshift record with a best-fitting redshift of $z=16.4$. This object is described in more detail in Section \ref{sec:z16}. 

\begin{figure*}
	\includegraphics[width=\textwidth]{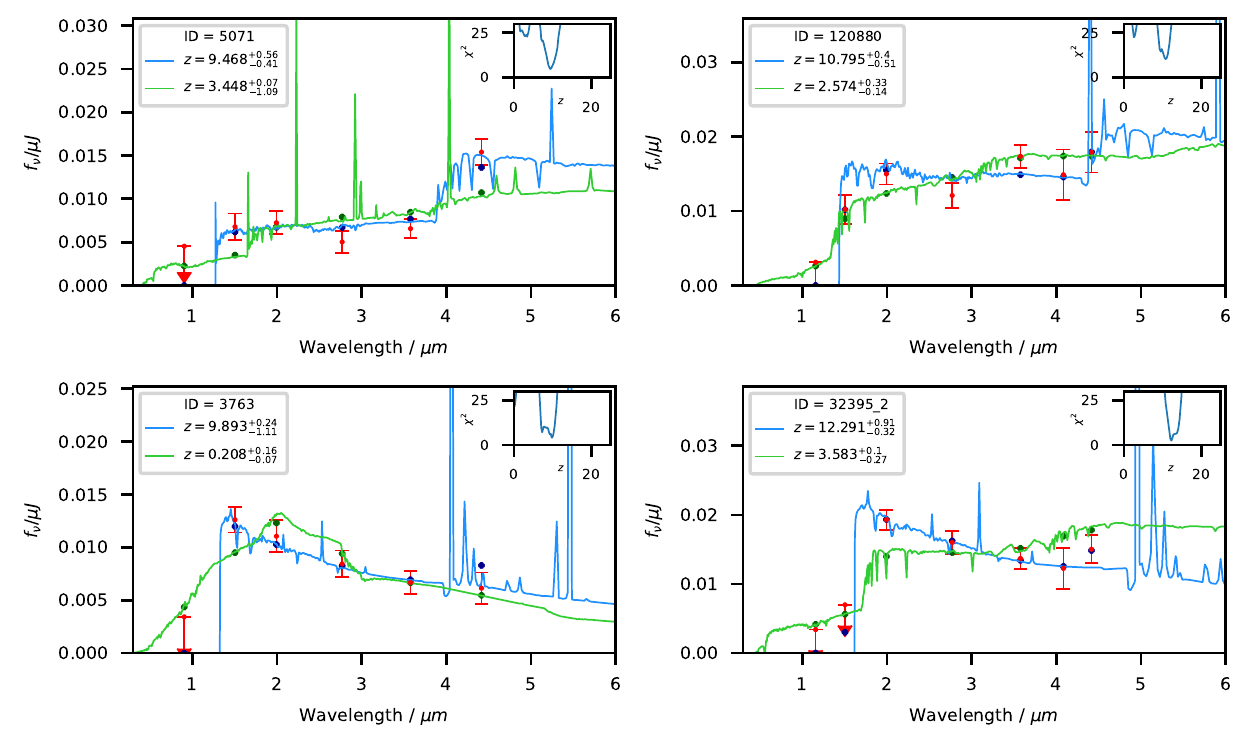}
    \caption{Spectral Energy Distribution (SED) fits for 4 example galaxies selected from within the final JWST high-redshift sample. The blue line shows the best-fitting (preferred) high-redshift solution, the green line shows the best-fitting (alternative) low-redshift solution, and the red points show the measured photometry (at 76$\%$ of total flux). The solid blue and green circles represent the model photometry of the best fitting high and low redshift templates respectively. The $\chi^2$ as a function of redshift is shown in the inset panels. These four galaxies have been chosen to be illustrative of the SEDs displayed by the galaxies found at redshifts $z\sim 9, 10, 11, 12$.}
    \label{fig:example_SEDs}
\end{figure*}

In Fig.\,\ref{fig:example_SEDs} we show illustrative examples of 
the SEDs of 4 of our JWST-selected galaxies, at redshifts $z \simeq 9, 10, 11$ \& 12.

The effective area available in which to search for high-redshift galaxies within each field was computed after masking the regions dominated by bright foreground sources, and removing areas of increased noise towards the edge of the imaging. 
The resulting effective area available for high-redshift galaxy detection/selection in each JWST field is listed in Table \ref{tab:area}.
These areas are also then used consistently in the calculation of the luminosity function in Section~\ref{sec:LF}.  We make a conservative estimate for the area in SMACS0723, including the removal of the highly-lensed region centred on the cluster. Only 1 of the 23 galaxies found in the SMACS0723 field lay within the excluded area, SMACS-34086, and therefore this is not included in the calculation of the luminosity function (note that this highly-lensed galaxy has been spectroscopically confirmed with NIRSpec at $z=8.948$; \citealt{carnall2022}). 

The UV magnitude of each source was determined and corrected to total assuming a point-source correction (see Section \ref{sec:LF}). We make further corrections to objects that look particularly extended by performing manual aperture photometry in 0.5" apertures as this is where the curve-of-growth in F200W looks approximately flat. 
The corrected absolute UV magnitudes for all the sources in our final $z > 8.5$ sample are given in Table \ref{tab:jwst_sample}, along with their positions and redshifts. Sources with extra corrections to total are marked with an asterisk.

\begin{table}
	\centering
	\caption{The derived effective areas available for robust high-redshift galaxy selection in each of three JWST fields used in this work.}
	\label{tab:area}
	\begin{tabular}{lc} 
		\hline
		Field & Area\\
		 & [arcmin$^2$]\\
		\hline
		CEERS & 31.7\\
		GLASS & 6.1\\
		SMACS0723 & 6.3\\
		\hline
	\end{tabular}
\end{table}

\begin{table}
	\centering
	\caption{The best-fitting photometric redshifts from \texttt{EAZY} for the final sample of high-redshift ($z > 8.5$) galaxies found in the combined JWST fields, ranked by photometric redshift. The first column gives the source ID, with $z_{\rm phot}$ for each object then presented in Column 2. Column 3 gives the derived rest-frame UV total magnitude of each galaxy. The coordinates for each source are given in the final two columns. We note here that sources 1698 and 17487 have also been independently discovered in the GLASS imaging by \citet{naidu2022} and \citet{castellano2022} and sources 34086, 44711 and 44566 have been spectroscopically confirmed at $z=8.498$, $z=7.663$ and $z=7.665$ respectively \citep{carnall2022}. The total UV magnitudes of sources indicated with an asterisk incorporate an additional correction to account for extended flux.}

	\label{tab:jwst_sample}
	\def\arraystretch{1.35}
	\begin{tabular}{lcccc} 
		\hline
		ID & $z_{\rm phot}$ & $M_{\rm UV}$ & RA & DEC\\
		\hline
43031 & 8.57$^{+0.65}_{-1.15}$ & $-18.43$ & 07:23:27.86 & $-$73:26:19.66 \\
29274\_4 & 8.86$^{+1.37}_{-0.60}$ & $-18.41$ & 14:19:27.39 & \phantom{$-$}52:51:46.90 \\
1434\_2 & 9.16$^{+1.30}_{-0.27}$ & $-18.82$ & 14:19:26.11 & \phantom{$-$}52:52:52.38 \\
44085 & 9.26$^{+0.20}_{-0.24}$ & $-18.25$ & 07:23:26.72 & $-$73:26:10.54 \\
38697 & 9.36$^{+0.38}_{-0.25}$ & $-18.86$ & 07:23:27.84 & $-$73:26:19.91 \\
5071 & 9.47$^{+0.56}_{-0.41}$ & $-18.02$ & 07:22:56.86 & $-$73:29:23.50 \\
44711$^*$ & 9.47$^{+0.24}_{-0.20}$ & $-20.14$ & 07:23:20.16 & $-$73:26:04.32 \\
43866 & 9.47$^{+0.38}_{-0.30}$ & $-18.14$ & 07:23:25.60 & $-$73:26:12.43 \\
34086 & 9.47$^{+0.27}_{-0.38}$ & $-17.87$ & 07:23:26.24 & $-$73:26:57.00 \\
14391 & 9.47$^{+0.36}_{-0.21}$ & $-18.81$ & 07:22:47.73 & $-$73:28:28.32 \\
12682 & 9.57$^{+0.50}_{-0.63}$ & $-18.95$ & 07:22:38.95 & $-$73:28:30.40 \\
44566$^*$ & 9.68$^{+0.14}_{-0.25}$ & $-20.68$ & 07:23:22.74 & $-$73:26:06.26 \\
22480 & 9.68$^{+0.36}_{-0.47}$ & $-18.50$ & 07:22:45.81 & $-$73:27:46.60 \\
15019 & 9.68$^{+0.12}_{-2.14}$ & $-18.67$ & 07:22:58.27 & $-$73:28:19.56 \\
12218 & 9.68$^{+0.59}_{-0.62}$ & $-19.28$ & 07:22:35.06 & $-$73:28:33.00 \\
3398 & 9.68$^{+0.10}_{-2.33}$ & $-18.21$ & 07:22:35.37 & $-$73:29:38.64 \\
6200 & 9.79$^{+0.24}_{-2.17}$ & $-18.52$ & 07:22:41.51 & $-$73:29:10.64 \\
7606 & 9.89$^{+0.20}_{-2.17}$ & $-18.08$ & 07:22:29.56 & $-$73:29:05.69 \\
3763 & 9.89$^{+0.24}_{-1.11}$ & $-18.99$ & 07:22:49.14 & $-$73:29:31.18 \\
1698$^*$ & 10.45$^{+0.26}_{-0.16}$ & $-20.62$ & 00:14:02.86 & $-$30:22:18.62 \\
20976\_4 & 10.45$^{+0.52}_{-0.96}$ & $-18.80$ & 14:19:36.30 & \phantom{$-$}52:50:49.18 \\
6647 & 10.45$^{+0.40}_{-0.92}$ & $-18.88$ & 14:19:14.67 & \phantom{$-$}52:48:49.76 \\
3710 & 10.45$^{+0.33}_{-0.68}$ & $-19.06$ & 14:19:24.03 & \phantom{$-$}52:48:28.98 \\
4063 & 10.45$^{+0.44}_{-2.45}$ & $-18.03$ & 07:22:52.31 & $-$73:29:32.39 \\
30585 & 10.56$^{+0.25}_{-0.52}$ & $-19.35$ & 14:19:35.34 & \phantom{$-$}52:50:37.87 \\
73150 & 10.56$^{+0.29}_{-1.10}$ & $-19.07$ & 14:19:26.78 & \phantom{$-$}52:54:16.59 \\
21071\_2 & 10.68$^{+0.28}_{-1.70}$ & $-19.27$ & 14:19:36.72 & \phantom{$-$}52:55:22.63 \\
20757 & 10.68$^{+0.49}_{-1.14}$ & $-17.88$ & 07:23:12.47 & $-$73:28:01.74 \\
6415 & 10.79$^{+0.45}_{-0.66}$ & $-19.13$ & 00:14:00.28 & $-$30:21:25.87 \\
120880 & 10.79$^{+0.40}_{-0.51}$ & $-19.43$ & 14:20:10.56 & \phantom{$-$}52:59:39.51 \\
26598 & 10.79$^{+0.26}_{-1.68}$ & $-18.47$ & 07:22:50.56 & $-$73:27:37.89 \\
61486 & 11.15$^{+0.37}_{-0.35}$ & $-19.61$ & 14:19:23.73 & \phantom{$-$}52:53:00.98 \\
622\_4 & 11.27$^{+0.48}_{-0.60}$ & $-18.92$ & 14:19:16.54 & \phantom{$-$}52:47:47.36 \\
33593\_2 & 11.27$^{+0.58}_{-0.28}$ & $-19.58$ & 14:19:37.59 & \phantom{$-$}52:56:43.82 \\
77241$^*$ & 11.27$^{+0.39}_{-0.70}$ & $-19.60$ & 14:19:41.47 & \phantom{$-$}52:54:41.49 \\
		\hline
	\end{tabular}
\end{table}

\begin{table}
\ContinuedFloat
	\centering
	\caption{Continued.}

	\label{tab:jwst_sample2}
	\def\arraystretch{1.35}
	\begin{tabular}{lcccc} 
		\hline
		ID & $z_{\rm phot}$ & $M_{\rm UV}$ & RA & DEC\\
		\hline
5268\_2 & 11.40$^{+0.30}_{-1.11}$ & $-19.16$ & 14:19:19.68 & \phantom{$-$}52:53:32.11 \\
127682 & 11.40$^{+0.59}_{-0.51}$ & $-19.07$ & 14:19:59.25 & \phantom{$-$}53:00:21.34 \\
26409\_4 & 11.90$^{+1.60}_{-0.70}$ & $-18.84$ & 14:19:38.48 & \phantom{$-$}52:51:18.12 \\
8347 & 11.90$^{+0.27}_{-0.39}$ & $-19.09$ & 07:22:56.36 & $-$73:29:00.52 \\
10566 & 12.03$^{+0.57}_{-0.26}$ & $-19.70$ & 07:23:03.47 & $-$73:28:46.99 \\
32395\_2 & 12.29$^{+0.91}_{-0.32}$ & $-19.89$ & 14:19:46.35 & \phantom{$-$}52:56:32.82 \\
1566 & 12.29$^{+1.50}_{-0.44}$ & $-18.77$ & 07:22:39.16 & $-$73:30:00.83 \\
17487 & 12.42$^{+0.27}_{-0.14}$ & $-20.89$ & 00:13:59.76 & $-$30:19:29.07 \\
27535\_4 & 12.56$^{+1.75}_{-0.27}$ & $-19.42$ & 14:19:27.31 & \phantom{$-$}52:51:29.23 \\
93316$^*$ & 16.39$^{+0.32}_{-0.22}$ & $-21.66$ & 14:19:39.49 & \phantom{$-$}52:56:34.94 \\

		\hline
	\end{tabular}
\end{table}

\section{The Luminosity Function}
\label{sec:LF}
Having completed the selection and refinement of our final galaxy samples at $z>7.5$ we proceed to compute the UV luminosity function at $z= 8, 9$ with a redshift bin width of $\Delta z=1$ and at $z=10.5$ with a bin width $\Delta z=2$ and $z=13.25$ with a bin width $\Delta z=3.5$. The UV absolute magnitude was estimated for each galaxy from the best-fitting SED template using a tophat filter centered on $\lambda_{\rm rest}~=~1500 \text{\AA}$ with a width of $100 \text{\AA}$. This was then converted to an absolute magnitude using
\begin{equation}
    M_{UV} = m_{1500} - 5 \log_{10} \left(\frac{D_{L}}{10}\right) + 2.5 \log_{10} \left(1+z\right),
    \label{eq:Muv}
\end{equation}
where $m_{1500}$ is the apparent magnitude at $\lambda_{\rm rest}=1500 \text{\AA}$, $D_{L}$ is luminosity distance in parsecs and $z$ is the best fitting redshift of the source.

\subsection{Determining completeness}
An accurate derivation of the UV LF requires a reliable estimate of how complete the samples are near to the magnitude limits of the imaging data. To calculate this we ran completeness simulations to estimate the fraction of galaxies we expect to recover as a function of observed magnitude in the detection images. This was done by injecting fake point sources (based on the measured PSF in the imaging) into three different regions of the different detection images. The sources were injected in steps of apparent aperture magnitude and the fraction of successfully recovered sources was measured at each step. This was performed $10$ times in each of the three regions and a median was taken of the resulting $30$ simulations. For COSMOS we treated the "Deep" and "Ultra-Deep" regions as separate fields. This produced the completeness as a function of apparent AB magnitude and was performed for each detection image in all of the fields analysed in this work. For GLASS and CEERS, 3 cutouts were made in random areas of the fields and sources were injected into each.  From F200W (AB) $= 24 - 31$ in steps of 0.1, 800 sources were injected into each area at each step. Therefore, a total of 336,000 sources were injected into both fields combined. For SMACS0793 there were 2 larger cutout areas with one covering the cluster field and one covering the parallel field with 1000 sources inserted in each area at each step. Therefore, a total of 140,000 sources were injected into SMACS0793. This was then implemented in the determination of the UV LF as described in Section \ref{sec:number_density}. 

\subsection{Determining number density}
\label{sec:number_density}
The binned co-moving number density of sources per absolute magnitude, $\Phi(M_{\rm UV})$, was determined using the $1/V_{\rm max}$ method \citep{schmidt1968}.
The equation for $\Phi(M_{\rm UV})$ is given by 
\begin{equation}
    \Phi(M_{\rm UV}) \Delta M = \sum_{i=1}^{N} \left( \frac{1}{C(m_{\rm AB}) V_{\rm max}} \right),
    \label{eq:LF}
\end{equation}
where $N$ is the number of galaxies in each bin, $C(m_{\rm AB})$ is the completeness as a function of $m_{\rm AB}$ in the detection image, and $V_{\rm max}$ is the maximum volume the galaxy could occupy and still be detected in the appropriate filter for the given redshift. This was determined by redshifting each galaxy from its measured photo-$z$ until it could no longer be detected in the appropriate detection filter (at a redshift $z_{\rm max}$). The volume, $V_{\rm max}$, is the difference in co-moving volume between the co-moving volume at $z_{\rm max}$ and at the minimum redshift for that sample (i.e. for the $z=8$ sample the minimum redshift is $z=7.5$). In the case of the SMACS0723 field, the volume was adjusted by the magnification factor which was computed from \texttt{GLAFIC} \citep{oguri2010}. As mentioned in Section \ref{sec:jwst_sample}, the SMACS0723 sources included in the LF calculation lie outside the most magnified region. Therefore the magnification values of the sources used in the LF calculation are $\lesssim2$.

The completeness factor, $C(m_{\rm AB})$, takes into account how incomplete the given galaxy sample is at the apparent AB magnitude in the detection image. This then leads to a value for the number density of galaxies in the given UV absolute magnitude bin. The Poisson uncertainties were calculated using the confidence intervals from \citet{gehrels1986}. 

\begin{figure*}
	\includegraphics[width=\textwidth]{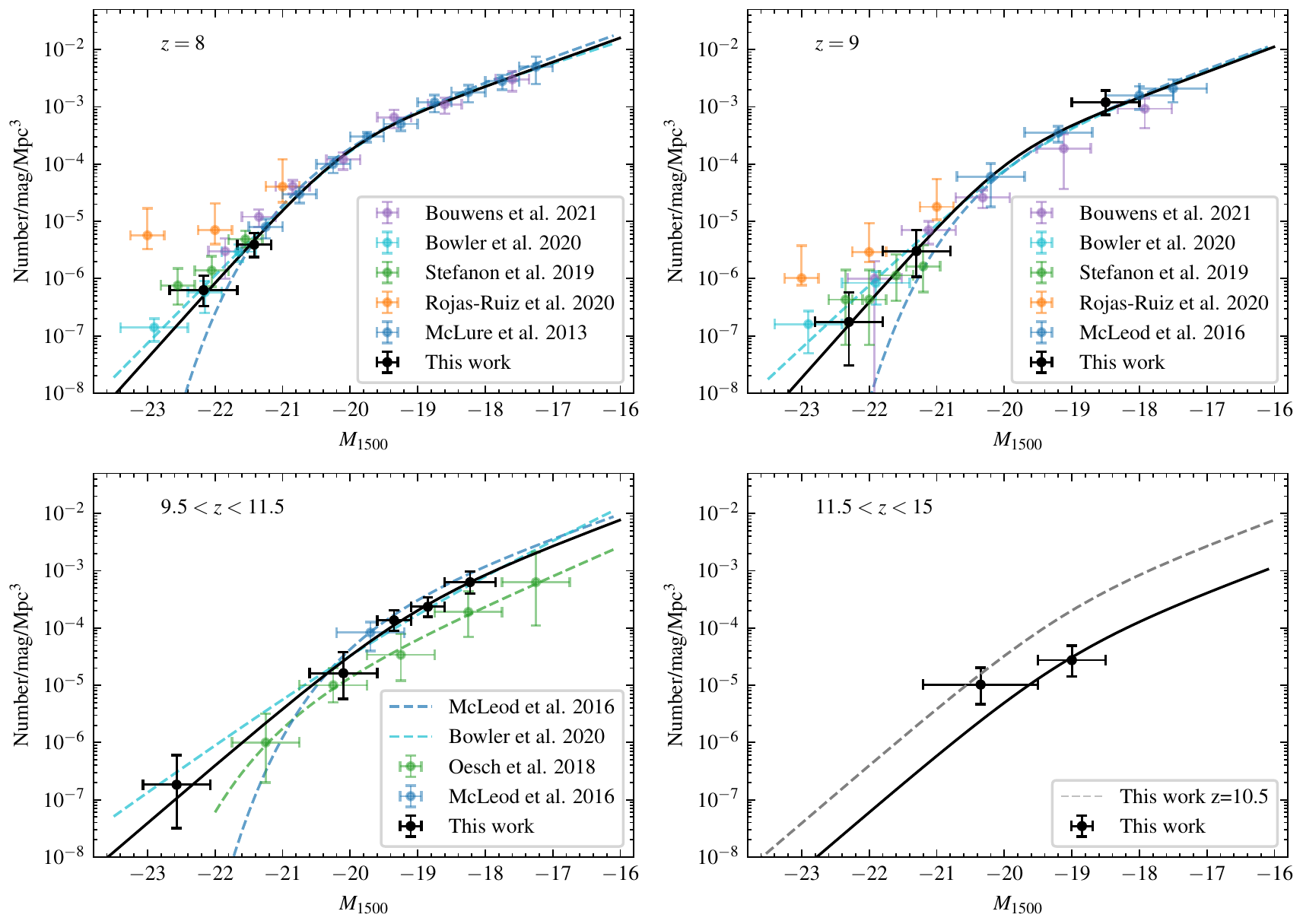}
    \caption{The rest-frame UV LF at $z=8, 9, 10.5$ and $z=13.25$ shown as black points. We include data points from \citet{mclure2013,mcleod2016,bouwens2021,oesch2018}. The best-fitting Schechter functions from \citet{mclure2013} and \citet{mcleod2016} are shown as the dashed blue lines at $z=8,9,10.5$. The best-fitting Schechter function from \citet{oesch2018} at $z=10$ is shown as the dashed green line. Our best-fitting double power laws are shown as solid black lines with the best-fitting double power laws from \citet{bowler2020} at $z=8,9,10.5$ shown as dashed cyan lines. }
    \label{fig:LF}
\end{figure*}

Our determinations of the UV LF at $z=8, 9, 10.5, 13.25$ are shown in Fig. \ref{fig:LF} and tabulated in Table \ref{tab:LF}. The combined dataset from the ground and space allows for increased dynamic range in our LF determination although this is only possible up to $z\sim11$ due to the limited wavelength range accessible from the ground. At $z=8$ the new data contributing to the LF is purely from the COSMOS sample and therefore we place new constraints on the bright end alone. Our results are in good agreement with \citet{bowler2020} and show a clear deviation from the \citet{mclure2013} Schechter function at the bright end. At $z=9$ our results at the bright end are in also good agreement with \citet{bowler2020} and our faint-end bins determined using the JWST sample are in good agreement with \citet{mcleod2016}. Our JWST results at $z=9$ are also in good agreement with \citet{bouwens2021}. At $z=10.5$ our new source from the UltraVISTA imaging allows us to compute a bin at the bright end of this LF and our JWST sample provides a fainter bin. The $z=13.25$ LF was calculated from the JWST sample alone as this redshift range cannot be probed from the ground. This shows a modest evolution from $z=9$ to $z=10.5$ and then further decline to $z=13.25$.

\begin{table}
	\centering
	\caption{Computed UV LF data points using the derived sample of $z>7.5$ galaxies from UltraVISTA and JWST. The columns show redshift, the central UV absolute magnitude of the bin, the bin width and the source number density within the bin, along with uncertainties.}
	\label{tab:LF}
	\def\arraystretch{1.35}
	\begin{tabular}{lccc} 
		\hline
		$z$ & $M_{\rm UV}$ & \phantom{4}$\Delta M$ & $\phi$\phantom{4444444}\\
		 & [mag] & [mag] & [$10^{-6}$/mag/Mpc$^{3}$]\\
		\hline
		8 & $-$22.17 & 1.0 & 0.63$^{+0.50}_{-0.30}$\\
		8 & $-$21.42 & 0.5 & 3.92$^{+2.34}_{-1.56}$\\
		\hline
		9 & $-$22.30 & 1.0 & 0.17$^{+0.40}_{-0.14}$\\
		9 & $-$21.30 & 1.0 & 3.02$^{3.98}_{-1.95}$\\
		9 & $-$18.50 & 1.0 & 1200$^{+717}_{-476}$\\
		\hline
		10.5 & $-$22.57 & 1.0 & 0.18$^{+0.42}_{-0.15}$\\
		10.5 & $-$20.10 & 1.0 & 16.2$^{+21.4}_{-10.5}$\\
		10.5 & $-$19.35 & 0.5 & 136.0$^{+67.2}_{-47.1}$\\
		10.5 & $-$18.85 & 0.5 & 234.9$^{+107}_{-76.8}$\\
		10.5 & $-$18.23 & 0.75 & 630.8$^{+340}_{-233}$\\
		\hline
		13.25 & $-$20.35 & 1.7 & 10.3$^{+9.98}_{-5.59}$\\
		13.25 & $-$19.00 & 1.0 & 27.4$^{+21.7}_{-13.1}$\\
		
		\hline
	\end{tabular}
\end{table}

\subsubsection{Luminosity function fitting}
As shown by this work, and the results of \citet{bowler2014,bowler2015,bowler2020}, a double-power law (DPL) function is a more suitable fit to the UV LF at $z\geq8$. Therefore, we fit a DPL to all the new high-redshift luminosity function data derived here. The fitting was done using the Scipy \citep[][]{virtanen2020} curve\_fit function which uses a least-squares method to fit the data. At $z=8$ and $z=9$ we combine our data points with the data points from \citet{mclure2013} and \citet{mcleod2016}, respectively. In these two redshift bins, our DPL fits are consistent with those previously derived by \citet{bowler2020}. At $z=9$ and $z=10.5$ we fix the faint-end slope to $\alpha=-2.10$ which is the value derived in \citet{bowler2020}. 
At $z>11.5$ we lose dynamic range due to the lack of ground-based objects and limited sample size. Therefore, for our fit at $z=13.25$ we also fix the bright-end slope to $\beta=-3.53$ and $M^*=-19.12$, the best-fitting values obtained at $z=10.5$, and allow only $\phi^*$ to vary as a free parameter. 
With these constraints, the best-fitting DPL function for each redshift is shown as the solid black line in Fig. \ref{fig:LF}. The best fitting parameters for our DPL fits are listed in Table \ref{tab:dpl_params}.

\begin{table}
	\centering
	\caption{The derived parameter values for the best-fitting double power-law (DPL) models fitted to our data over the redshift range $8 < z < 15$. The LF fits derived at $z=8$ and $z=9$ utilised the new data presented here along with the data-points presented by \citet{mclure2013} and \citet{mcleod2016}. At higher redshifts the fits are based purely on the new analysis and galaxy samples presented in this work. The first column gives the central redshift of the binned LF. This is followed by the values of the best-fitting characteristic density $\phi^*$, the best-fitting or fixed characteristic absolute magnitude $M^*$, the fitted or assumed faint-end slope $\alpha$, and the fitted or adopted bright-end slope $\beta$ (see text for details). In the case where a parameter was fixed, the value is denoted with an asterisk.}
	\label{tab:dpl_params}
    \setlength{\tabcolsep}{4pt} 
	\renewcommand{\arraystretch}{1.15} 
	\begin{tabular}{lcccc} 
		\hline
		$z$ & $\phi^*$ & $M^*$ & $\alpha$ & $\beta$\\
		 & [$10^{-4}$/mag/Mpc$^{3}$] & [mag] &  & \\
		\hline
		8 & $3.30\pm3.41$ & $-20.02\pm0.55$ & $-2.04\pm$0.29 & $-4.26\pm$0.50\\
		9 & $2.10\pm1.68$ & $-19.93\pm0.58$ & $-2.10^*$ & $-4.29\pm$0.69\\
		10.5 & $3.32\pm8.96$ & $-19.12\pm1.68$ & $-2.10^*$ & $-3.53\pm$1.06\\
		13.25 & $0.51\pm0.22$ & $-19.12^*$ & $-2.10^*$ & $-3.53^*$\\

		\hline
	\end{tabular}
\end{table}

\subsubsection{Comparison to the results of HST pure parallel imaging}
There have been many recent attempts to use pure-parallel imaging with HST to try to determine the bright end of the galaxy UV luminosity function at high redshifts. To illustrate this we over-plot in Fig.\,2 the LF points produced by  \citet{rojasruiz2020}, which are in close agreement with the results reported by  \citet{leethochawalit2022} (see also \citealt{bagley2022}), derived from SuperBORG, the largest-area  HST pure-parallel survey covering $\simeq 1000$\,arcmin$^2$. However, it is well known that the limited wavelength coverage available in much/most of the HST pure-parallel imaging makes high-redshift galaxy samples derived from surveys such as SuperBORG extremely vulnerable to contamination. It is thus perhaps unsurprising that their derived number densities are completely inconsistent with our estimates of the bright end of the UV LF. For example, assuming the number density of the brightest bin from \citet{rojasruiz2020} at $z=8$, we would have expected to find $\sim 50$ galaxies in our brightest luminosity bin, whereas in fact we only find 4 galaxies. At $z=9$ the results from HST pure parallel studies are also inconsistent with our findings at the bright end of the LF. At $M_{UV}=-22.3$, assuming the number density from \citet{rojasruiz2020}, we should have found $\sim21$ galaxies and assuming the number densities from \citet{finkelstein2022b} and \citet{bagley2022} we should have detected $\sim17$ galaxies. In this study we only find 1 galaxy in this bin.

\subsection{The cosmic SFRD at $\mathbf {z \geq 8}$}
\label{sec:sfrd}
\begin{figure}
	\includegraphics[width=\columnwidth]{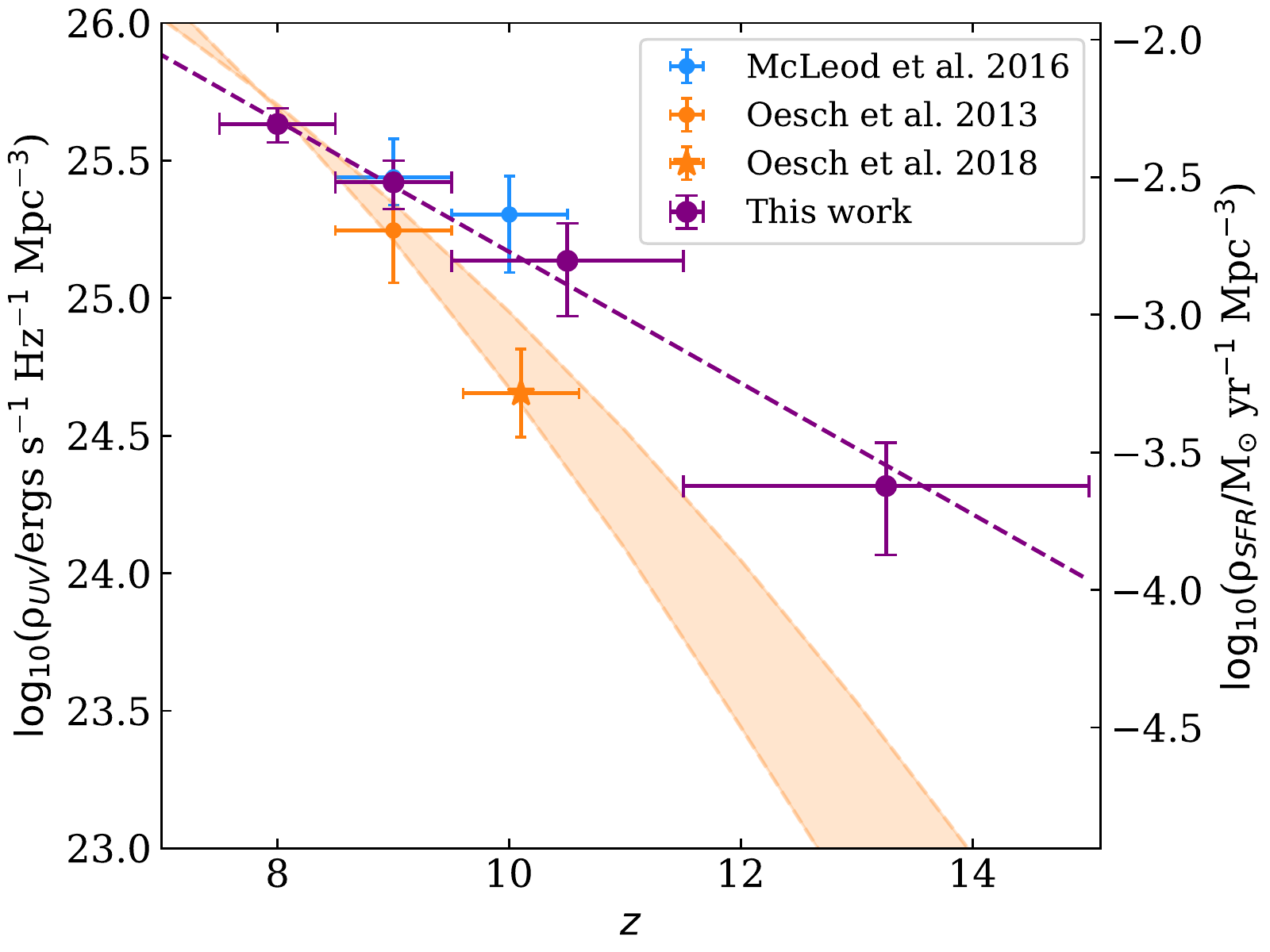}
    \caption{The redshift evolution of the UV luminosity density $\rho_{\rm UV}$ and therefore the cosmic star-formation rate density $\rho_{\rm SFR}$ at $z>7$ with our new measurements at $z=8$, $z=9$, $z=10.5$, and $z=13.25$ (purple circular data points). Estimates at $z\simeq9-10$ from \citet{oesch2013,oesch2018} and \citet{mcleod2016} are shown by the orange and blue data points respectively. All values were determined using a limit of $M_{\rm UV}=-17$ in the luminosity-weighted integral. The dashed purple line shows a log-linear fit to our data points with the solution ${\rm{log}_{10}(\rho_{\rm UV}) = (-0.231\pm0.037) z + (27.5\pm 0.3)}$. The shaded orange region shows the halo evolution model from \citet{oesch2018}. In contrast to the \citet{oesch2018} claim of a rapid fall-off in $\rho_{\rm UV}$ at $z>8$, our data favour a steady, exponential decline in $\rho_{\rm UV}$ up to $z\simeq15$ \citep[consistent with the result of][]{mcleod2016}.}
    \label{fig:sfrd}
\end{figure}

The evolution of the UV luminosity density and cosmic star-formation rate density at $z>8$ has been a point of contention in recent HST-based studies, with \citet{oesch2018} concluding in favour of a rapid decline at $z>8$, whereas  \citet{mcleod2016} presented evidence for a much smoother, gradual decline extending out to higher redshifts. Using our new estimates of the evolving UV LF at $z=8, 9, 10.5$ \& 13.25, we perform a luminosity-weighted integral of our best-fitting double-power law fits to determine the evolution of UV luminosity density, $\rho_{\rm UV}$. We integrate down to $M_{\rm UV}=-17$ and use the same limit to integrate the LFs from \citet{oesch2014,oesch2018} and \citet{mcleod2016}. The UV luminosity density is converted to the cosmic star-formation rate $\rho_{\rm SFR}$ using the conversion factor $\cal{K}$$_{\rm UV} = 1.15 \times 10^{-28}$ M$_{\odot}$ yr$^{-1}$/erg s$^{-1}$ Hz$^{-1}$ \citep{madau2014}.  The results are shown in Fig. \ref{fig:sfrd}. We also perform a log-linear fit to our data points (motivated, in part, by the analytical work of \citet{hernquist2003}) and find that the evolution of $\rho_{\rm UV}$ with redshift is well described by: 
\begin{equation}
\rm{log}_{10}(\rho_{\rm UV}) = (-0.231\pm0.037) z + (27.5\pm 0.3).
\end{equation}
We also plot a rapidly descending halo evolution model from \citet{oesch2018}. This is shown as the shaded blue region. Our results are inconsistent with this function.

\begin{figure*}
	\includegraphics[width=\textwidth]{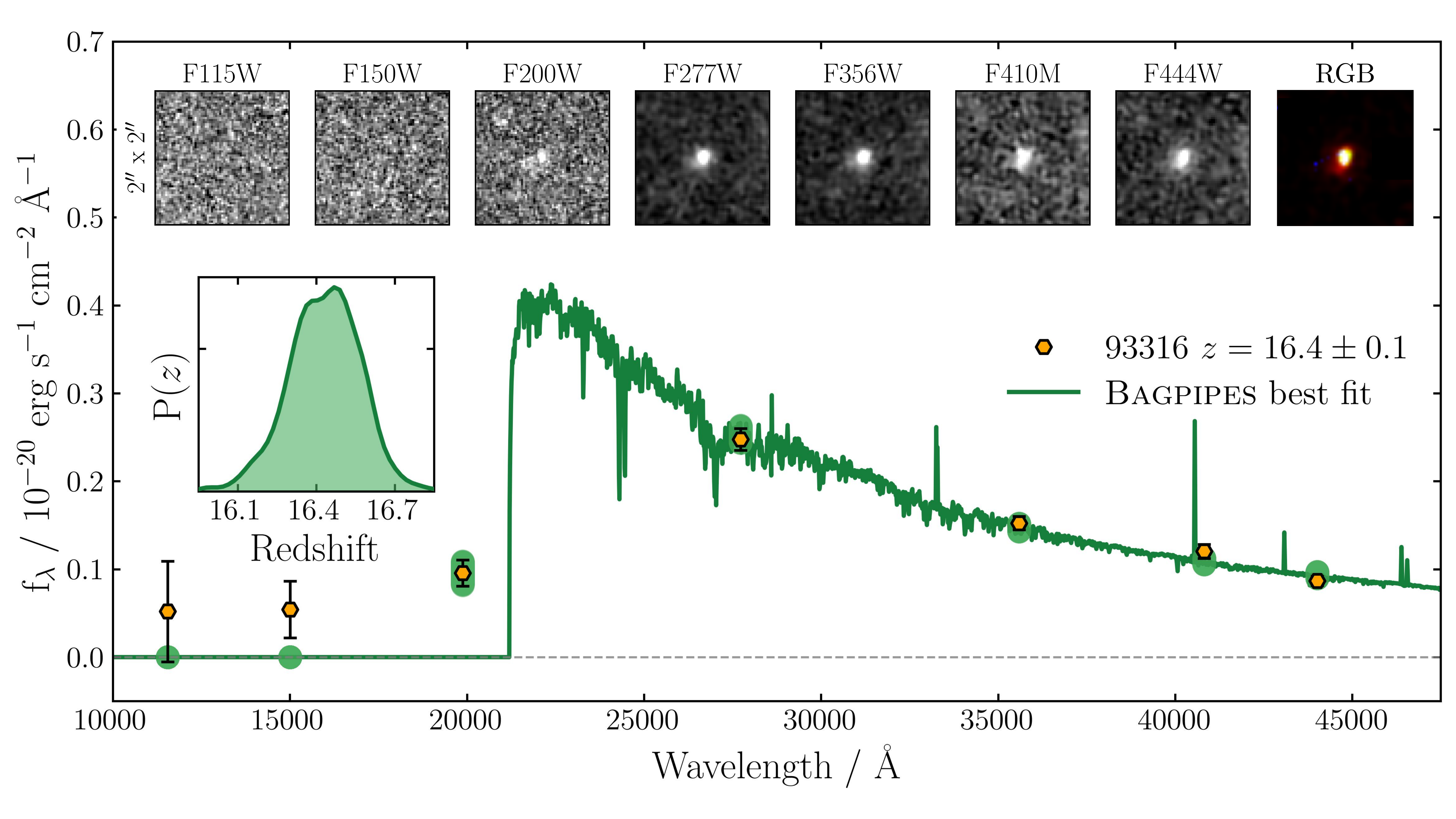}
    \caption{The highest-redshift object in our sample, CEERS-93316. The NIRCam photometric measurements are plotted in the SED plot as golden hexagons, while $2^{\prime\prime}\times2^{\prime\prime}$ postage-stamp images in each band are shown above the SED. The \texttt{Bagpipes} model we fit in Section \ref{sec:z16} is shown in green. The posterior distribution for redshift is shown in the inset panel, which is centred on $z=16.4$, and is fully consistent with the value of $z=16.4$ quoted in Table \ref{tab:jwst_sample} from \texttt{EAZY}. The fortuitous positioning of the F200W and F277W bands relative to the Lyman break allows such a precise redshift estimate. The rest-frame near-UV slope, $\beta=-2.06\pm0.25$ indicates no evidence for an unusual (i.e. Population III dominated) stellar population. The galaxy has a stellar mass of log$_{10}(M_*/$M$_\odot) = 9.0^{+0.4}_{-0.5}$}.
    \label{fig:93316}
\end{figure*}

\section{Discussion}
\label{sec:discussion}

\subsection{The early growth of galaxies and cosmic star-formation}
The results of this study provide dramatic, early confirmation of the long-anticipated power of JWST to chart the evolution of galaxies back to within $< 300$\,Myr of the Big Bang. In addition to independently uncovering the 2 bright galaxies at redshifts $z \simeq 10.5$ and $z \simeq 12.5$ recently reported by \citet{naidu2022} and \citet{castellano2022} from the GLASS NIRCam imaging, we here report the discovery of an additional 43 galaxies at $ z > 8.5$ from the combined SMACS0723+GLASS+CEERS ERO/ERS JWST NIRCam imaging, covering a total effective area of $\simeq 44$\, arcmin$^2$.

This analysis also reaffirms the importance of large dynamic range (in galaxy luminosity) for properly constraining the form of the evolving galaxy luminosity function (LF), here provided by the addition of the $\simeq 2$\,deg$^2$ of deep near-infrared imaging now delivered by UltraVISTA DR5. Here we report the discovery of 16 bright galaxies in the redshift range $7.5 < z < 10.5$ from this relatively wide-area ground-based imaging: providing crucial information on the bright end of the galaxy LF, at least out to $z \simeq 10$. To fill in the luminosity/volume gap between UltraVISTA and the early small-area JWST surveys, to extend the study of brighter/rarer objects to $z \ge 10$, and to improve the statistical robustness of our results, forthcoming larger-area deep JWST imaging surveys, such as PRIMER (GO 1837)\footnote{https://primer-jwst.github.io/}, will be key.

Encouragingly, even with the existing ground-based+JWST dataset we are able to draw a number of firm conclusions. First, we are able to settle the long-standing uncertainty/dispute over the evolution of UV luminosity density, $\rho_{\rm UV}$ (and hence star formation rate density, $\rho_{\rm SFR}$) at redshifts beyond $z \simeq 8$. Contrary to the conclusion reached by \citet{oesch2014,oesch2018} that $\rho_{\rm UV}$ falls off rapidly at $z > 8$, the results shown in Fig. \ref{fig:sfrd} support the conclusion of \citet{mcleod2015,mcleod2016} (in line with the long-standing analytical prediction of 
\citet{hernquist2003}), that $\rho_{\rm UV}$ continues to display a steady, exponential decline with increasing redshift out to at least $z \simeq 12$. Clearly, galaxy formation commenced at even higher redshifts ($z > 15$).

Second, armed with the UltraVISTA DR5 and JWST ERO/ERS imaging, we are able to reaffirm the findings of \citet{bowler2020} that, at the highest redshifts ($z \ge 7$) the galaxy LF can no longer be adequately described by a Schechter function, but instead evolves into a more gently declining double power-law, a functional form that more closely mirrors the shape of the underlying dark-matter halo function. As described in \citet{bowler2020}, and then discussed further in \citet{adams2022}, this is arguably as expected, as we look back into an era when neither mass-quenching, nor significant dust-obscuration are able to curtail the luminosities of the brightest galaxies.  

Third, while less surprising, and hence undoubtedly less important, it is worth noting that we can completely rule out the very high number densities of bright high-redshift galaxies reported from recent pure-parallel HST studies \citep[e.g.][]{rojasruiz2020}. If the conclusions of such studies were valid, we would have detected an order-of-magnitude more bright galaxies at $z > 7$ in UltraVISTA than were actually revealed by the data. There are several obvious lessons here, including a timely reminder of the importance of insisting on a sufficient number of photometric bands to properly constrain galaxy photometric redshifts and basic physical properties.

\subsection{A galaxy candidate at $\mathbf{z = 16.4}$}
\label{sec:z16}
Finally, in addition to the other sources discussed earlier in this work, we highlight the apparent discovery of an object with a well-constrained photometric redshift of $z=16.4$, corresponding to a time just $\simeq250$\,Myr after the Big Bang. This object was selected as a F150W dropout in the CEERS imaging data. However, it is much brighter in F277W and the longer-wavelength filters than in F200W, indicating that the Lyman break lies towards the red end of the F200W filter. This fortuitous alignment of filters produces a particularly well-constrained photometric redshift, with all three of the photometric redshift codes discussed above returning consistent values at $z=16.3-16.5$ and no plausible secondary low-redshift solutions.

Since our first-reported discovery of this extreme redshift source, some follow-up studies have discussed the plausibility of an alternative, lower-redshift solution. In particular,  \citet{naidu2022b} propose a extreme emission line solution from a passive or dusty star-forming galaxy. They argue this solution becomes plausible from SED fitting when arbitrarily boosting the uncertainties on the fluxes to 20\% due to zero-point uncertainty. However, as discussed here, the robustness of the high redshift solution is unchanged with updated zero-points. Secondly as described in \citet{naidu2022b}, the extreme, arguably unphysical nature of the $z \sim 5$ solution and very limited redshift range required to reproduce the photometry, imply that the $z\sim5$ solution is still less plausible than the $z\sim16$ solution. This is consistent with our secondary solution from \texttt{EAZY} which is at $z=4.9$ but, even with the revised NIRCam photometric zeropoints, still has a much higher $\chi^2$ than the best fitting $z=16.4$ solution ($\Delta \chi^2=20.3$).

The object is also clearly resolved in the NIRCam imaging data, and so cannot be a low-mass star or unobscured active galactic nucleus. We have re-calculated the  photometry for this object using a variety of aperture sizes, but this does not change our recovered redshift. Having searched extensively, we are currently unable to find any plausible explanation for this object, other than a galaxy at a new redshift record of $z=16.4$.

In order to constrain the physical properties of this galaxy, we fit our photometric data using the \texttt{Bagpipes} spectral fitting code \citep{carnall2018}. We use the same configuration described in \cite{Carnall2020, carnall2022}, including the 2016 updated version of the \cite{bruzual2003} stellar population models with the MILES stellar spectral library, an emission line prescription calculated using the \texttt{Cloudy} photoionization code \citep{Ferland2017}, the \cite{Salim2018} dust attenuation model and a constant SFH model. The time before observation at which stars began forming is varied from 1\,Myr to the age of the Universe with a logarithmic prior.

The results of our spectral fitting analysis are shown in Fig.\,\ref{fig:93316}. We obtain a photometric redshift of $z=16.4\pm0.1$, in good agreement with the other three codes discussed above. We also measure a stellar mass, log$_{10}(M_*/$M$_\odot) = 9.0^{+0.4}_{-0.5}$ (assuming a \cite{Kroupa2001} initial mass function), with the large uncertainty due to the lack of rest-frame optical data. We recover log$_{10}($SFR/$\rm{M}_{\odot}$yr$^{-1}) = 1.0^{+0.3}_{-0.5}$ and a (mass-weighted) mean stellar age of $20^{+40}_{-10}$ Myr. Assuming our constant SFH model, we find that star formation first began in this object between 120 and 220 Myr after the Big Bang ($z=18-26$). 

By separate analysis, we recover a rest-frame UV spectral slope, $\beta=-2.06\pm0.25$. In combination with our \texttt{Bagpipes} fit finding dust attenuation, $A_\mathrm{V}$, consistent with zero, this suggests no evidence for an unusual (i.e., Population III dominated) stellar population.

An important consideration is whether this new, relatively massive galaxy at such an extreme redshift is consistent with the $\Lambda-$CDM halo-mass function. We consider this object in the context of the analysis presented by \cite{behroozi2018}, which provides cumulative number density thresholds for high-redshift galaxies in $\Lambda-$CDM under the assumption that all gas available to halos is converted into stars. Across our survey volume of $\simeq10^{5}$ Mpc$^3$ from $15 < z < 17$ we find that our object falls close to, but does not significantly exceed the $\Lambda-$CDM limit calculated by \cite{behroozi2018}. Since initial publication of this work as a preprint, this finding has subsequently been validated by several other authors \citep{Boylan-Kolchin2022, Lovell2022}.

These comparisons with the halo-mass function are strongly dependent on our stellar-mass estimate of log$_{10}(M_*/$M$_\odot) = 9.0^{+0.4}_{-0.5}$, which assumes a \cite{Kroupa2001} IMF. At high redshift, some theoretical and observational evidence points towards a more top-heavy IMF (e.g. \citealt{Sneppen2022}), which would lead to a reduction in our implied stellar mass, moving our object further from the $\Lambda-$CDM limit.

Another potential issue that could impact our stellar mass estimate is binary stellar evolution. This extends the lifetimes of very massive stars, increasing the luminosity of young stellar populations (e.g. \citealt{Eldridge2017}), potentially resulting in a reduction of our stellar mass estimate. As an additional test, we re-fit this galaxy with \texttt{bagpipes} using the Binary Population and Spectral Synthesis (BPASS) models (v2.2.1; \citealt{Stanway2018}), including binary evolution, with an upper IMF mass limit of 300\,M$_\odot$. Under the assumption of these alternative stellar models, we recover a $\simeq0.2$ dex lower stellar mass of log$_{10}(M_*/$M$_\odot) =8.8^{+0.4}_{-0.4}$.

\section{Conclusions}
\label{sec:conclusions}
We have re-reduced and analysed the early public \textit{James Webb Space Telescope} (JWST) ERO and ERS NIRCam imaging (SMACS0723, GLASS, CEERS) in combination with the latest deep ground-based near-infrared imaging in the COSMOS field provided by UltraVISTA DR5, with the aim of producing a new sample of galaxies at $z > 7.5$ to probe early galaxy evolution. Through careful galaxy candidate selection, and the use of a range of photometric redshift codes, we have assembled a combined sample of 61 high-redshift galaxies, 47 of which are reported here for the first time.

We have exploited this new sample, in tandem with pre-existing results from HST, 
to produce a new measurement of the evolving galaxy UV luminosity function (LF) over the redshift range $z = 8 - 15$. The luminosity-weighted integral of the evolving LF then yields a new estimate of the evolution of UV luminosity density ($\rho_{\rm UV}$), which we then convert into an estimate of declining cosmic star-formation rate density ($\rho_{\rm SFR}$) out to within $< 300$\, Myr of the Big Bang.

Our results confirm that the high-redshift LF evolves into a form that is best described by a double power-law (rather than a Schechter) function (at least up to $z\sim10$ as shown by the COSMOS analysis), and that the LF and the resulting derived $\rho_{\rm UV}$ (and thus $\rho_{\rm SFR}$), continues to decline gradually and steadily over this redshift range (as anticipated from previous studies which analysed the pre-existing data in a consistent manner). 

We provide details of the 61 high-redshift galaxy candidates, with full photometry, SED fits, and multi-band postage-stamp images presented in Appendices A and B. Our sample contains 6 galaxies at $z \ge 12$, one of which is the galaxy at $z = 12.4$ independently reported by \citet{naidu2022} and \citet{castellano2022}. However, the most distant object is one which appears to set a new redshift record as an apparently robust galaxy candidate at $z \simeq 16.4$. Given the apparently extreme nature of this source, we consider its physical properties and plausibility in detail.

The advances presented here emphasize the importance of achieving high dynamic range in studies of early galaxy evolution, and re-affirm the enormous potential of forthcoming larger JWST programmes to transform our understanding of the young Universe.

\section*{Acknowledgements}
 C. T. Donnan, D. J. McLeod, R. J. McLure, J. S. Dunlop, R. Begley, F. Cullen and M. L. Hamadouche acknowledge the support of the Science and Technology Facilities Council. A.C. Carnall thanks the Leverhulme Trust for their support via the Leverhulme Early Career Fellowship scheme. The Cosmic Dawn Center is funded by the Danish National Research Foundation under grant no.\,140.
 
Based on observations collected at the European Southern Observatory under ESO programme ID 179.A-2005 and 198.A-2003 and on data products produced by CALET and the Cambridge Astronomy Survey Unit on behalf of the UltraVISTA consortium.

This work is based [in part] on observations made with the NASA/ESA/CSA James Webb Space Telescope. The data were obtained from the Mikulski Archive for Space Telescopes at the Space Telescope Science Institute, which is operated by the Association of Universities for Research in Astronomy, Inc., under NASA contract NAS 5-03127 for JWST. These observations are associated with programs 1324, 1345, 2736. The Early Release Observations and associated materials were developed, executed, and compiled by the ERO production team:  Hannah Braun, Claire Blome, Matthew Brown, Margaret Carruthers, Dan Coe, Joseph DePasquale, Nestor Espinoza, Macarena Garcia Marin, Karl Gordon, Alaina Henry, Leah Hustak, Andi James, Ann Jenkins, Anton Koekemoer, Stephanie LaMassa, David Law, Alexandra Lockwood, Amaya Moro-Martin, Susan Mullally, Alyssa Pagan, Dani Player, Klaus Pontoppidan, Charles Proffitt, Christine Pulliam, Leah Ramsay, Swara Ravindranath, Neill Reid, Massimo Robberto, Elena Sabbi, Leonardo Ubeda. The EROs were also made possible by the foundational efforts and support from the JWST instruments, STScI planning and scheduling, and Data Management teams.

For the purpose of open access, the author has applied a Creative Commons Attribution (CC BY) licence to any Author Accepted Manuscript version arising from this submission.

\section*{Data Availability}

All JWST and HST data products are available via the Mikulski Archive for Space Telescopes (\url{https://mast.stsci.edu}). UltraVISTA DR5 will shortly be made available through ESO. Additional data products are available from the authors upon reasonable request.

\bibliographystyle{mnras}
\bibliography{jwst_paper} 

\begin{thebibliography}{}
\makeatletter
\relax
\def\mn@urlcharsother{\let\do\@makeother \do\$\do\&\do\#\do\^\do\_\do\%\do\~}
\def\mn@doi{\begingroup\mn@urlcharsother \@ifnextchar [ {\mn@doi@}
  {\mn@doi@[]}}
\def\mn@doi@[#1]#2{\def\@tempa{#1}\ifx\@tempa\@empty \href
  {http://dx.doi.org/#2} {doi:#2}\else \href {http://dx.doi.org/#2} {#1}\fi
  \endgroup}
\def\mn@eprint#1#2{\mn@eprint@#1:#2::\@nil}
\def\mn@eprint@arXiv#1{\href {http://arxiv.org/abs/#1} {{\tt arXiv:#1}}}
\def\mn@eprint@dblp#1{\href {http://dblp.uni-trier.de/rec/bibtex/#1.xml}
  {dblp:#1}}
\def\mn@eprint@#1:#2:#3:#4\@nil{\def\@tempa {#1}\def\@tempb {#2}\def\@tempc
  {#3}\ifx \@tempc \@empty \let \@tempc \@tempb \let \@tempb \@tempa \fi \ifx
  \@tempb \@empty \def\@tempb {arXiv}\fi \@ifundefined
  {mn@eprint@\@tempb}{\@tempb:\@tempc}{\expandafter \expandafter \csname
  mn@eprint@\@tempb\endcsname \expandafter{\@tempc}}}

\bibitem[\protect\citeauthoryear{{Adams}, {Bowler}, {Jarvis}, {Varadaraj}  \&
  {H{\"a}u{\ss}ler}}{{Adams} et~al.}{2022}]{adams2022}
{Adams} N.~J.,  {Bowler} R.~A.~A.,  {Jarvis} M.~J.,  {Varadaraj} R.~G.,
  {H{\"a}u{\ss}ler} B.,  2022, arXiv e-prints, \href
  {https://ui.adsabs.harvard.edu/abs/2022arXiv220709342A} {p. arXiv:2207.09342}

\bibitem[\protect\citeauthoryear{{Aihara} et~al.,}{{Aihara}
  et~al.}{2019}]{aihara2019}
{Aihara} H.,  et~al., 2019, \mn@doi [\pasj] {10.1093/pasj/psz103}, \href
  {https://ui.adsabs.harvard.edu/abs/2019PASJ...71..114A} {71, 114}

\bibitem[\protect\citeauthoryear{{Aird}, {Coil}, {Georgakakis}, {Nandra},
  {Barro}  \& {P{\'e}rez-Gonz{\'a}lez}}{{Aird} et~al.}{2015}]{aird2015}
{Aird} J.,  {Coil} A.~L.,  {Georgakakis} A.,  {Nandra} K.,  {Barro} G.,
  {P{\'e}rez-Gonz{\'a}lez} P.~G.,  2015, \mn@doi [\mnras]
  {10.1093/mnras/stv1062}, \href
  {https://ui.adsabs.harvard.edu/abs/2015MNRAS.451.1892A} {451, 1892}

\bibitem[\protect\citeauthoryear{{Arnouts}, {Cristiani}, {Moscardini},
  {Matarrese}, {Lucchin}, {Fontana}  \& {Giallongo}}{{Arnouts}
  et~al.}{1999}]{arnouts1999}
{Arnouts} S.,  {Cristiani} S.,  {Moscardini} L.,  {Matarrese} S.,  {Lucchin}
  F.,  {Fontana} A.,   {Giallongo} E.,  1999, \mn@doi [\mnras]
  {10.1046/j.1365-8711.1999.02978.x}, \href
  {https://ui.adsabs.harvard.edu/abs/1999MNRAS.310..540A} {310, 540}

\bibitem[\protect\citeauthoryear{{Bagley} et~al.,}{{Bagley}
  et~al.}{2022}]{bagley2022}
{Bagley} M.~B.,  et~al., 2022, arXiv e-prints, \href
  {https://ui.adsabs.harvard.edu/abs/2022arXiv220512980B} {p. arXiv:2205.12980}

\bibitem[\protect\citeauthoryear{{Behroozi} \& {Silk}}{{Behroozi} \&
  {Silk}}{2018}]{behroozi2018}
{Behroozi} P.,  {Silk} J.,  2018, \mn@doi [\mnras] {10.1093/mnras/sty945},
  \href {https://ui.adsabs.harvard.edu/abs/2018MNRAS.477.5382B} {477, 5382}

\bibitem[\protect\citeauthoryear{{Bertin} \& {Arnouts}}{{Bertin} \&
  {Arnouts}}{1996}]{bertin1996}
{Bertin} E.,  {Arnouts} S.,  1996, \mn@doi [\aaps] {10.1051/aas:1996164}, \href
  {https://ui.adsabs.harvard.edu/abs/1996A&AS..117..393B} {117, 393}

\bibitem[\protect\citeauthoryear{{Bouwens} et~al.,}{{Bouwens}
  et~al.}{2021}]{bouwens2021}
{Bouwens} R.~J.,  et~al., 2021, \mn@doi [\aj] {10.3847/1538-3881/abf83e}, \href
  {https://ui.adsabs.harvard.edu/abs/2021AJ....162...47B} {162, 47}

\bibitem[\protect\citeauthoryear{{Bouwens}, {Illingworth}, {Ellis}, {Oesch}  \&
  {Stefanon}}{{Bouwens} et~al.}{2022}]{bouwens2022}
{Bouwens} R.~J.,  {Illingworth} G.~D.,  {Ellis} R.~S.,  {Oesch} P.~A.,
  {Stefanon} M.,  2022, arXiv e-prints, \href
  {https://ui.adsabs.harvard.edu/abs/2022arXiv220511526B} {p. arXiv:2205.11526}

\bibitem[\protect\citeauthoryear{{Bowler} et~al.,}{{Bowler}
  et~al.}{2014}]{bowler2014}
{Bowler} R.~A.~A.,  et~al., 2014, \mn@doi [\mnras] {10.1093/mnras/stu449},
  \href {https://ui.adsabs.harvard.edu/abs/2014MNRAS.440.2810B} {440, 2810}

\bibitem[\protect\citeauthoryear{{Bowler} et~al.,}{{Bowler}
  et~al.}{2015}]{bowler2015}
{Bowler} R.~A.~A.,  et~al., 2015, \mn@doi [\mnras] {10.1093/mnras/stv1403},
  \href {https://ui.adsabs.harvard.edu/abs/2015MNRAS.452.1817B} {452, 1817}

\bibitem[\protect\citeauthoryear{{Bowler}, {Dunlop}, {McLure}  \&
  {McLeod}}{{Bowler} et~al.}{2017}]{bowler2017}
{Bowler} R.~A.~A.,  {Dunlop} J.~S.,  {McLure} R.~J.,   {McLeod} D.~J.,  2017,
  \mn@doi [\mnras] {10.1093/mnras/stw3296}, \href
  {https://ui.adsabs.harvard.edu/abs/2017MNRAS.466.3612B} {466, 3612}

\bibitem[\protect\citeauthoryear{{Bowler}, {Jarvis}, {Dunlop}, {McLure},
  {McLeod}, {Adams}, {Milvang-Jensen}  \& {McCracken}}{{Bowler}
  et~al.}{2020}]{bowler2020}
{Bowler} R.~A.~A.,  {Jarvis} M.~J.,  {Dunlop} J.~S.,  {McLure} R.~J.,  {McLeod}
  D.~J.,  {Adams} N.~J.,  {Milvang-Jensen} B.,   {McCracken} H.~J.,  2020,
  \mn@doi [\mnras] {10.1093/mnras/staa313}, \href
  {https://ui.adsabs.harvard.edu/abs/2020MNRAS.493.2059B} {493, 2059}

\bibitem[\protect\citeauthoryear{{Boylan-Kolchin}}{{Boylan-Kolchin}}{2022}]{Boylan-Kolchin2022}
{Boylan-Kolchin} M.,  2022, arXiv e-prints, \href
  {https://ui.adsabs.harvard.edu/abs/2022arXiv220801611B} {p. arXiv:2208.01611}

\bibitem[\protect\citeauthoryear{{Brammer}, {van Dokkum}  \& {Coppi}}{{Brammer}
  et~al.}{2008}]{brammer2008}
{Brammer} G.~B.,  {van Dokkum} P.~G.,   {Coppi} P.,  2008, \mn@doi [\apj]
  {10.1086/591786}, \href
  {https://ui.adsabs.harvard.edu/abs/2008ApJ...686.1503B} {686, 1503}

\bibitem[\protect\citeauthoryear{{Bruzual} \& {Charlot}}{{Bruzual} \&
  {Charlot}}{2003}]{bruzual2003}
{Bruzual} G.,  {Charlot} S.,  2003, \mn@doi [\mnras]
  {10.1046/j.1365-8711.2003.06897.x}, \href
  {https://ui.adsabs.harvard.edu/abs/2003MNRAS.344.1000B} {344, 1000}

\bibitem[\protect\citeauthoryear{{Carnall}, {McLure}, {Dunlop}  \&
  {Dav{\'e}}}{{Carnall} et~al.}{2018}]{carnall2018}
{Carnall} A.~C.,  {McLure} R.~J.,  {Dunlop} J.~S.,   {Dav{\'e}} R.,  2018,
  \mn@doi [\mnras] {10.1093/mnras/sty2169}, \href
  {http://adsabs.harvard.edu/abs/2018MNRAS.480.4379C} {480, 4379}

\bibitem[\protect\citeauthoryear{{Carnall} et~al.,}{{Carnall}
  et~al.}{2020}]{Carnall2020}
{Carnall} A.~C.,  et~al., 2020, \mn@doi [\mnras] {10.1093/mnras/staa1535},
  \href {https://ui.adsabs.harvard.edu/abs/2020MNRAS.496..695C} {496, 695}

\bibitem[\protect\citeauthoryear{{Carnall} et~al.,}{{Carnall}
  et~al.}{2022}]{carnall2022}
{Carnall} A.~C.,  et~al., 2022, arXiv e-prints, \href
  {https://ui.adsabs.harvard.edu/abs/2022arXiv220708778C} {p. arXiv:2207.08778}

\bibitem[\protect\citeauthoryear{{Castellano} et~al.,}{{Castellano}
  et~al.}{2022}]{castellano2022}
{Castellano} M.,  et~al., 2022, arXiv e-prints, \href
  {https://ui.adsabs.harvard.edu/abs/2022arXiv220709436C} {p. arXiv:2207.09436}

\bibitem[\protect\citeauthoryear{{Cooper} et~al.,}{{Cooper}
  et~al.}{2012}]{cooper2012}
{Cooper} M.~C.,  et~al., 2012, \mn@doi [\mnras]
  {10.1111/j.1365-2966.2011.19938.x}, \href
  {https://ui.adsabs.harvard.edu/abs/2012MNRAS.419.3018C} {419, 3018}

\bibitem[\protect\citeauthoryear{{Dunlop}}{{Dunlop}}{2013}]{dunlop2013}
{Dunlop} J.~S.,  2013, in {Wiklind} T.,  {Mobasher} B.,   {Bromm} V.,  eds,
  Astrophysics and Space Science Library Vol. 396, The First Galaxies. p.~223
  (\mn@eprint {arXiv} {1205.1543}), \mn@doi{10.1007/978-3-642-32362-1\_5}

\bibitem[\protect\citeauthoryear{{Eldridge}, {Stanway}, {Xiao}, {McClelland},
  {Taylor}, {Ng}, {Greis}  \& {Bray}}{{Eldridge} et~al.}{2017}]{Eldridge2017}
{Eldridge} J.~J.,  {Stanway} E.~R.,  {Xiao} L.,  {McClelland} L.~A.~S.,
  {Taylor} G.,  {Ng} M.,  {Greis} S.~M.~L.,   {Bray} J.~C.,  2017, \mn@doi
  [\pasa] {10.1017/pasa.2017.51}, \href
  {https://ui.adsabs.harvard.edu/abs/2017PASA...34...58E} {34, e058}

\bibitem[\protect\citeauthoryear{{Ellis} et~al.,}{{Ellis}
  et~al.}{2013}]{ellis2013}
{Ellis} R.~S.,  et~al., 2013, \mn@doi [\apjl] {10.1088/2041-8205/763/1/L7},
  \href {https://ui.adsabs.harvard.edu/abs/2013ApJ...763L...7E} {763, L7}

\bibitem[\protect\citeauthoryear{{Euclid Collaboration} et~al.,}{{Euclid
  Collaboration} et~al.}{2022}]{moneti2022}
{Euclid Collaboration} et~al., 2022, \mn@doi [\aap]
  {10.1051/0004-6361/202142361}, \href
  {https://ui.adsabs.harvard.edu/abs/2022A&A...658A.126E} {658, A126}

\bibitem[\protect\citeauthoryear{{Ferland} et~al.,}{{Ferland}
  et~al.}{2017}]{Ferland2017}
{Ferland} G.~J.,  et~al., 2017, \rmxaa, \href
  {https://ui.adsabs.harvard.edu/abs/2017RMxAA..53..385F} {53, 385}

\bibitem[\protect\citeauthoryear{{Finkelstein} et~al.,}{{Finkelstein}
  et~al.}{2015}]{finkelstein2015}
{Finkelstein} S.~L.,  et~al., 2015, \mn@doi [\apj]
  {10.1088/0004-637X/810/1/71}, \href
  {https://ui.adsabs.harvard.edu/abs/2015ApJ...810...71F} {810, 71}

\bibitem[\protect\citeauthoryear{{Finkelstein} et~al.,}{{Finkelstein}
  et~al.}{2019}]{finkelstein2019}
{Finkelstein} S.~L.,  et~al., 2019, \mn@doi [\apj] {10.3847/1538-4357/ab1ea8},
  \href {https://ui.adsabs.harvard.edu/abs/2019ApJ...879...36F} {879, 36}

\bibitem[\protect\citeauthoryear{{Finkelstein} et~al.,}{{Finkelstein}
  et~al.}{2022a}]{finkelstein2022}
{Finkelstein} S.~L.,  et~al., 2022a, arXiv e-prints, \href
  {https://ui.adsabs.harvard.edu/abs/2022arXiv220712474F} {p. arXiv:2207.12474}

\bibitem[\protect\citeauthoryear{{Finkelstein} et~al.,}{{Finkelstein}
  et~al.}{2022b}]{finkelstein2022b}
{Finkelstein} S.~L.,  et~al., 2022b, \mn@doi [\apj] {10.3847/1538-4357/ac3aed},
  \href {https://ui.adsabs.harvard.edu/abs/2022ApJ...928...52F} {928, 52}

\bibitem[\protect\citeauthoryear{{Gehrels}}{{Gehrels}}{1986}]{gehrels1986}
{Gehrels} N.,  1986, \mn@doi [\apj] {10.1086/164079}, \href
  {https://ui.adsabs.harvard.edu/abs/1986ApJ...303..336G} {303, 336}

\bibitem[\protect\citeauthoryear{{Grogin} et~al.,}{{Grogin}
  et~al.}{2011}]{grogin2011}
{Grogin} N.~A.,  et~al., 2011, \mn@doi [\apjs] {10.1088/0067-0049/197/2/35},
  \href {https://ui.adsabs.harvard.edu/abs/2011ApJS..197...35G} {197, 35}

\bibitem[\protect\citeauthoryear{{Hernquist} \& {Springel}}{{Hernquist} \&
  {Springel}}{2003}]{hernquist2003}
{Hernquist} L.,  {Springel} V.,  2003, \mn@doi [\mnras]
  {10.1046/j.1365-8711.2003.06499.x}, \href
  {https://ui.adsabs.harvard.edu/abs/2003MNRAS.341.1253H} {341, 1253}

\bibitem[\protect\citeauthoryear{{Hudelot} et~al.,}{{Hudelot}
  et~al.}{2012}]{hudelot2012}
{Hudelot} P.,  et~al., 2012, VizieR Online Data Catalog, \href
  {https://ui.adsabs.harvard.edu/abs/2012yCat.2317....0H} {p. II/317}

\bibitem[\protect\citeauthoryear{{Ilbert} et~al.,}{{Ilbert}
  et~al.}{2006}]{ilbert2006}
{Ilbert} O.,  et~al., 2006, \mn@doi [\aap] {10.1051/0004-6361:20065138}, \href
  {https://ui.adsabs.harvard.edu/abs/2006A&A...457..841I} {457, 841}

\bibitem[\protect\citeauthoryear{{Kroupa}}{{Kroupa}}{2001}]{Kroupa2001}
{Kroupa} P.,  2001, \mn@doi [\mnras] {10.1046/j.1365-8711.2001.04022.x}, \href
  {https://ui.adsabs.harvard.edu/abs/2001MNRAS.322..231K} {322, 231}

\bibitem[\protect\citeauthoryear{{Lawrence} et~al.,}{{Lawrence}
  et~al.}{2007}]{lawrence2007}
{Lawrence} A.,  et~al., 2007, \mn@doi [\mnras]
  {10.1111/j.1365-2966.2007.12040.x}, \href
  {https://ui.adsabs.harvard.edu/abs/2007MNRAS.379.1599L} {379, 1599}

\bibitem[\protect\citeauthoryear{{Leethochawalit}, {Roberts-Borsani},
  {Morishita}, {Trenti}  \& {Treu}}{{Leethochawalit}
  et~al.}{2022a}]{leethochawalit2022}
{Leethochawalit} N.,  {Roberts-Borsani} G.,  {Morishita} T.,  {Trenti} M.,
  {Treu} T.,  2022a, arXiv e-prints, \href
  {https://ui.adsabs.harvard.edu/abs/2022arXiv220515388L} {p. arXiv:2205.15388}

\bibitem[\protect\citeauthoryear{{Leethochawalit} et~al.,}{{Leethochawalit}
  et~al.}{2022b}]{leethochawalit2022b}
{Leethochawalit} N.,  et~al., 2022b, arXiv e-prints, \href
  {https://ui.adsabs.harvard.edu/abs/2022arXiv220711135L} {p. arXiv:2207.11135}

\bibitem[\protect\citeauthoryear{{Lovell}, {Harrison}, {Harikane}, {Tacchella}
  \& {Wilkins}}{{Lovell} et~al.}{2022}]{Lovell2022}
{Lovell} C.~C.,  {Harrison} I.,  {Harikane} Y.,  {Tacchella} S.,   {Wilkins}
  S.~M.,  2022, arXiv e-prints, \href
  {https://ui.adsabs.harvard.edu/abs/2022arXiv220810479L} {p. arXiv:2208.10479}

\bibitem[\protect\citeauthoryear{{Madau} \& {Dickinson}}{{Madau} \&
  {Dickinson}}{2014}]{madau2014}
{Madau} P.,  {Dickinson} M.,  2014, \mn@doi [\araa]
  {10.1146/annurev-astro-081811-125615}, \href
  {https://ui.adsabs.harvard.edu/abs/2014ARA&A..52..415M} {52, 415}

\bibitem[\protect\citeauthoryear{{McCracken} et~al.,}{{McCracken}
  et~al.}{2012}]{mccracken2012}
{McCracken} H.~J.,  et~al., 2012, \mn@doi [\aap] {10.1051/0004-6361/201219507},
  \href {https://ui.adsabs.harvard.edu/abs/2012A&A...544A.156M} {544, A156}

\bibitem[\protect\citeauthoryear{{McLeod}, {McLure}, {Dunlop}, {Robertson},
  {Ellis}  \& {Targett}}{{McLeod} et~al.}{2015}]{mcleod2015}
{McLeod} D.~J.,  {McLure} R.~J.,  {Dunlop} J.~S.,  {Robertson} B.~E.,  {Ellis}
  R.~S.,   {Targett} T.~A.,  2015, \mn@doi [\mnras] {10.1093/mnras/stv780},
  \href {https://ui.adsabs.harvard.edu/abs/2015MNRAS.450.3032M} {450, 3032}

\bibitem[\protect\citeauthoryear{{McLeod}, {McLure}  \& {Dunlop}}{{McLeod}
  et~al.}{2016}]{mcleod2016}
{McLeod} D.~J.,  {McLure} R.~J.,   {Dunlop} J.~S.,  2016, \mn@doi [\mnras]
  {10.1093/mnras/stw904}, \href
  {https://ui.adsabs.harvard.edu/abs/2016MNRAS.459.3812M} {459, 3812}

\bibitem[\protect\citeauthoryear{{McLure} et~al.,}{{McLure}
  et~al.}{2011}]{mclure2011}
{McLure} R.~J.,  et~al., 2011, \mn@doi [\mnras]
  {10.1111/j.1365-2966.2011.19626.x}, \href
  {https://ui.adsabs.harvard.edu/abs/2011MNRAS.418.2074M} {418, 2074}

\bibitem[\protect\citeauthoryear{{McLure} et~al.,}{{McLure}
  et~al.}{2013}]{mclure2013}
{McLure} R.~J.,  et~al., 2013, \mn@doi [\mnras] {10.1093/mnras/stt627}, \href
  {https://ui.adsabs.harvard.edu/abs/2013MNRAS.432.2696M} {432, 2696}

\bibitem[\protect\citeauthoryear{{Merlin} et~al.,}{{Merlin}
  et~al.}{2015}]{merlin2015}
{Merlin} E.,  et~al., 2015, \mn@doi [\aap] {10.1051/0004-6361/201526471}, \href
  {https://ui.adsabs.harvard.edu/abs/2015A&A...582A..15M} {582, A15}

\bibitem[\protect\citeauthoryear{{Naidu} et~al.,}{{Naidu}
  et~al.}{2022a}]{naidu2022}
{Naidu} R.~P.,  et~al., 2022a, arXiv e-prints, \href
  {https://ui.adsabs.harvard.edu/abs/2022arXiv220709434N} {p. arXiv:2207.09434}

\bibitem[\protect\citeauthoryear{{Naidu} et~al.,}{{Naidu}
  et~al.}{2022b}]{naidu2022b}
{Naidu} R.~P.,  et~al., 2022b, arXiv e-prints, \href
  {https://ui.adsabs.harvard.edu/abs/2022arXiv220802794N} {p. arXiv:2208.02794}

\bibitem[\protect\citeauthoryear{{Newman} et~al.,}{{Newman}
  et~al.}{2013}]{newman2013}
{Newman} J.~A.,  et~al., 2013, \mn@doi [\apjs] {10.1088/0067-0049/208/1/5},
  \href {https://ui.adsabs.harvard.edu/abs/2013ApJS..208....5N} {208, 5}

\bibitem[\protect\citeauthoryear{{Oesch} et~al.,}{{Oesch}
  et~al.}{2013}]{oesch2013}
{Oesch} P.~A.,  et~al., 2013, \mn@doi [\apj] {10.1088/0004-637X/773/1/75},
  \href {https://ui.adsabs.harvard.edu/abs/2013ApJ...773...75O} {773, 75}

\bibitem[\protect\citeauthoryear{{Oesch} et~al.,}{{Oesch}
  et~al.}{2014}]{oesch2014}
{Oesch} P.~A.,  et~al., 2014, \mn@doi [\apj] {10.1088/0004-637X/786/2/108},
  \href {https://ui.adsabs.harvard.edu/abs/2014ApJ...786..108O} {786, 108}

\bibitem[\protect\citeauthoryear{{Oesch}, {Bouwens}, {Illingworth}, {Labb{\'e}}
   \& {Stefanon}}{{Oesch} et~al.}{2018}]{oesch2018}
{Oesch} P.~A.,  {Bouwens} R.~J.,  {Illingworth} G.~D.,  {Labb{\'e}} I.,
  {Stefanon} M.,  2018, \mn@doi [\apj] {10.3847/1538-4357/aab03f}, \href
  {https://ui.adsabs.harvard.edu/abs/2018ApJ...855..105O} {855, 105}

\bibitem[\protect\citeauthoryear{{Oguri}}{{Oguri}}{2010}]{oguri2010}
{Oguri} M.,  2010, \mn@doi [\pasj] {10.1093/pasj/62.4.1017}, \href
  {https://ui.adsabs.harvard.edu/abs/2010PASJ...62.1017O} {62, 1017}

\bibitem[\protect\citeauthoryear{{Oke}}{{Oke}}{1974}]{oke1974}
{Oke} J.~B.,  1974, \mn@doi [\apjs] {10.1086/190287}, \href
  {https://ui.adsabs.harvard.edu/abs/1974ApJS...27...21O} {27, 21}

\bibitem[\protect\citeauthoryear{{Oke} \& {Gunn}}{{Oke} \&
  {Gunn}}{1983}]{oke1983}
{Oke} J.~B.,  {Gunn} J.~E.,  1983, \mn@doi [\apj] {10.1086/160817}, \href
  {https://ui.adsabs.harvard.edu/abs/1983ApJ...266..713O} {266, 713}

\bibitem[\protect\citeauthoryear{{Planck Collaboration} et~al.,}{{Planck
  Collaboration} et~al.}{2020}]{planck2020}
{Planck Collaboration} et~al., 2020, \mn@doi [\aap]
  {10.1051/0004-6361/201833910}, \href
  {https://ui.adsabs.harvard.edu/abs/2020A&A...641A...6P} {641, A6}

\bibitem[\protect\citeauthoryear{{Pontoppidan} et~al.,}{{Pontoppidan}
  et~al.}{2022}]{pontoppidan2022}
{Pontoppidan} K.~M.,  et~al., 2022, \mn@doi [\apjl] {10.3847/2041-8213/ac8a4e},
  \href {https://ui.adsabs.harvard.edu/abs/2022ApJ...936L..14P} {936, L14}

\bibitem[\protect\citeauthoryear{{Rigby} et~al.,}{{Rigby}
  et~al.}{2022}]{rigby2022}
{Rigby} J.,  et~al., 2022, arXiv e-prints, \href
  {https://ui.adsabs.harvard.edu/abs/2022arXiv220705632R} {p. arXiv:2207.05632}

\bibitem[\protect\citeauthoryear{{Robertson}}{{Robertson}}{2021}]{robertson2021}
{Robertson} B.~E.,  2021, arXiv e-prints, \href
  {https://ui.adsabs.harvard.edu/abs/2021arXiv211013160R} {p. arXiv:2110.13160}

\bibitem[\protect\citeauthoryear{{Robertson}, {Ellis}, {Furlanetto}  \&
  {Dunlop}}{{Robertson} et~al.}{2015}]{robertson2015}
{Robertson} B.~E.,  {Ellis} R.~S.,  {Furlanetto} S.~R.,   {Dunlop} J.~S.,
  2015, \mn@doi [\apjl] {10.1088/2041-8205/802/2/L19}, \href
  {https://ui.adsabs.harvard.edu/abs/2015ApJ...802L..19R} {802, L19}

\bibitem[\protect\citeauthoryear{{Rojas-Ruiz}, {Finkelstein}, {Bagley},
  {Stevans}, {Finkelstein}, {Larson}, {Mechtley}  \& {Diekmann}}{{Rojas-Ruiz}
  et~al.}{2020}]{rojasruiz2020}
{Rojas-Ruiz} S.,  {Finkelstein} S.~L.,  {Bagley} M.~B.,  {Stevans} M.,
  {Finkelstein} K.~D.,  {Larson} R.,  {Mechtley} M.,   {Diekmann} J.,  2020,
  \mn@doi [\apj] {10.3847/1538-4357/ab7659}, \href
  {https://ui.adsabs.harvard.edu/abs/2020ApJ...891..146R} {891, 146}

\bibitem[\protect\citeauthoryear{{Salim}, {Boquien}  \& {Lee}}{{Salim}
  et~al.}{2018}]{Salim2018}
{Salim} S.,  {Boquien} M.,   {Lee} J.~C.,  2018, \mn@doi [\apj]
  {10.3847/1538-4357/aabf3c}, \href
  {https://ui.adsabs.harvard.edu/abs/2018ApJ...859...11S} {859, 11}

\bibitem[\protect\citeauthoryear{{Schmidt}}{{Schmidt}}{1968}]{schmidt1968}
{Schmidt} M.,  1968, \mn@doi [\apj] {10.1086/149446}, \href
  {https://ui.adsabs.harvard.edu/abs/1968ApJ...151..393S} {151, 393}

\bibitem[\protect\citeauthoryear{{Sneppen}, {Steinhardt}, {Hensley}, {Jermyn},
  {Mostafa}  \& {Weaver}}{{Sneppen} et~al.}{2022}]{Sneppen2022}
{Sneppen} A.,  {Steinhardt} C.~L.,  {Hensley} H.,  {Jermyn} A.~S.,  {Mostafa}
  B.,   {Weaver} J.~R.,  2022, \mn@doi [\apj] {10.3847/1538-4357/ac695e}, \href
  {https://ui.adsabs.harvard.edu/abs/2022ApJ...931...57S} {931, 57}

\bibitem[\protect\citeauthoryear{{Stanway} \& {Eldridge}}{{Stanway} \&
  {Eldridge}}{2018}]{Stanway2018}
{Stanway} E.~R.,  {Eldridge} J.~J.,  2018, \mn@doi [\mnras]
  {10.1093/mnras/sty1353}, \href
  {https://ui.adsabs.harvard.edu/abs/2018MNRAS.479...75S} {479, 75}

\bibitem[\protect\citeauthoryear{{Stark}}{{Stark}}{2016}]{stark2016}
{Stark} D.~P.,  2016, \mn@doi [\araa] {10.1146/annurev-astro-081915-023417},
  \href {https://ui.adsabs.harvard.edu/abs/2016ARA&A..54..761S} {54, 761}

\bibitem[\protect\citeauthoryear{{Stefanon} et~al.,}{{Stefanon}
  et~al.}{2019}]{stefanon2019}
{Stefanon} M.,  et~al., 2019, \mn@doi [\apj] {10.3847/1538-4357/ab3792}, \href
  {https://ui.adsabs.harvard.edu/abs/2019ApJ...883...99S} {883, 99}

\bibitem[\protect\citeauthoryear{{Treu} et~al.,}{{Treu}
  et~al.}{2022}]{treu2022}
{Treu} T.,  et~al., 2022, \mn@doi [\apj] {10.3847/1538-4357/ac8158}, \href
  {https://ui.adsabs.harvard.edu/abs/2022ApJ...935..110T} {935, 110}

\bibitem[\protect\citeauthoryear{{Virtanen} et~al.,}{{Virtanen}
  et~al.}{2020}]{virtanen2020}
{Virtanen} P.,  et~al., 2020, {scipy/scipy: SciPy 1.5.3}, Zenodo,
  \mn@doi{10.5281/zenodo.4100507}

\bibitem[\protect\citeauthoryear{{Weaver} et~al.,}{{Weaver}
  et~al.}{2022}]{weaver2022}
{Weaver} J.~R.,  et~al., 2022, \mn@doi [\apjs] {10.3847/1538-4365/ac3078},
  \href {https://ui.adsabs.harvard.edu/abs/2022ApJS..258...11W} {258, 11}

\makeatother
\end{thebibliography}

\appendix

\section{SEDs and postage-stamp images}
The SEDs and postage-stamp images of all 55 sources are provided in this Appendix. In Figs.\,\ref{fig:COSMOS_SEDs_1}-\ref{fig:COSMOS_SEDs_2} we show the best-fitting SEDs for the 16 COSMOS galaxy candidates. In Figs.\,\ref{fig:jwst_SEDs1}-\ref{fig:jwst_SEDs5} we show the best-fitting SEDs for the 45 JWST-selected galaxy candidates. The postage-stamp images for the galaxies in the COSMOS sample are presented in Figs.\,\ref{fig:COSMOS_cuts_1}-\ref{fig:COSMOS_cuts_2}. The postage-stamp images of the JWST-selected galaxies are shown in Figs.\,\ref{fig:jwst_cuts_1}-\ref{fig:jwst_cuts_6}.

\begin{figure*}
	\includegraphics[width=\textwidth]{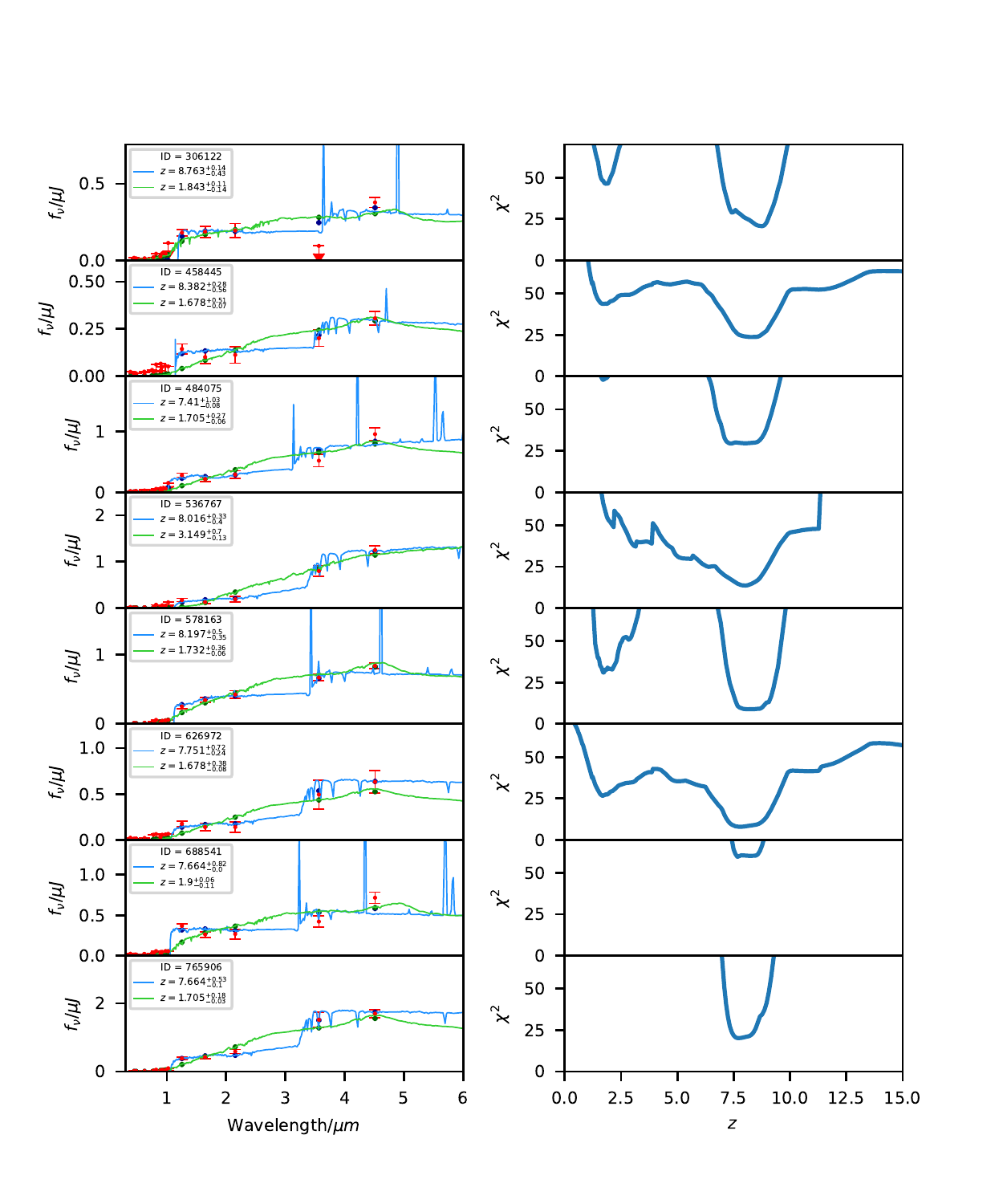}
    \caption{The measured photometry and best-fitting SED from \texttt{EAZY} for the $z>7.5$ galaxies found the COSMOS/UltraVISTA field. For each object the best-fitting high-redshift solution is shown in blue with the best-fitting alternative low-redshift solution plotted in green. Non-detections at the $2\sigma$ level are shown as downward arrows. The solid blue and green circles represent the model photometry of the best fitting high and low redshift templates respectively.} The panels in the right-hand column show $\chi^2$ as a function of redshift for each object.
    \label{fig:COSMOS_SEDs_1}
\end{figure*}

\begin{figure*}
	\includegraphics[width=\textwidth]{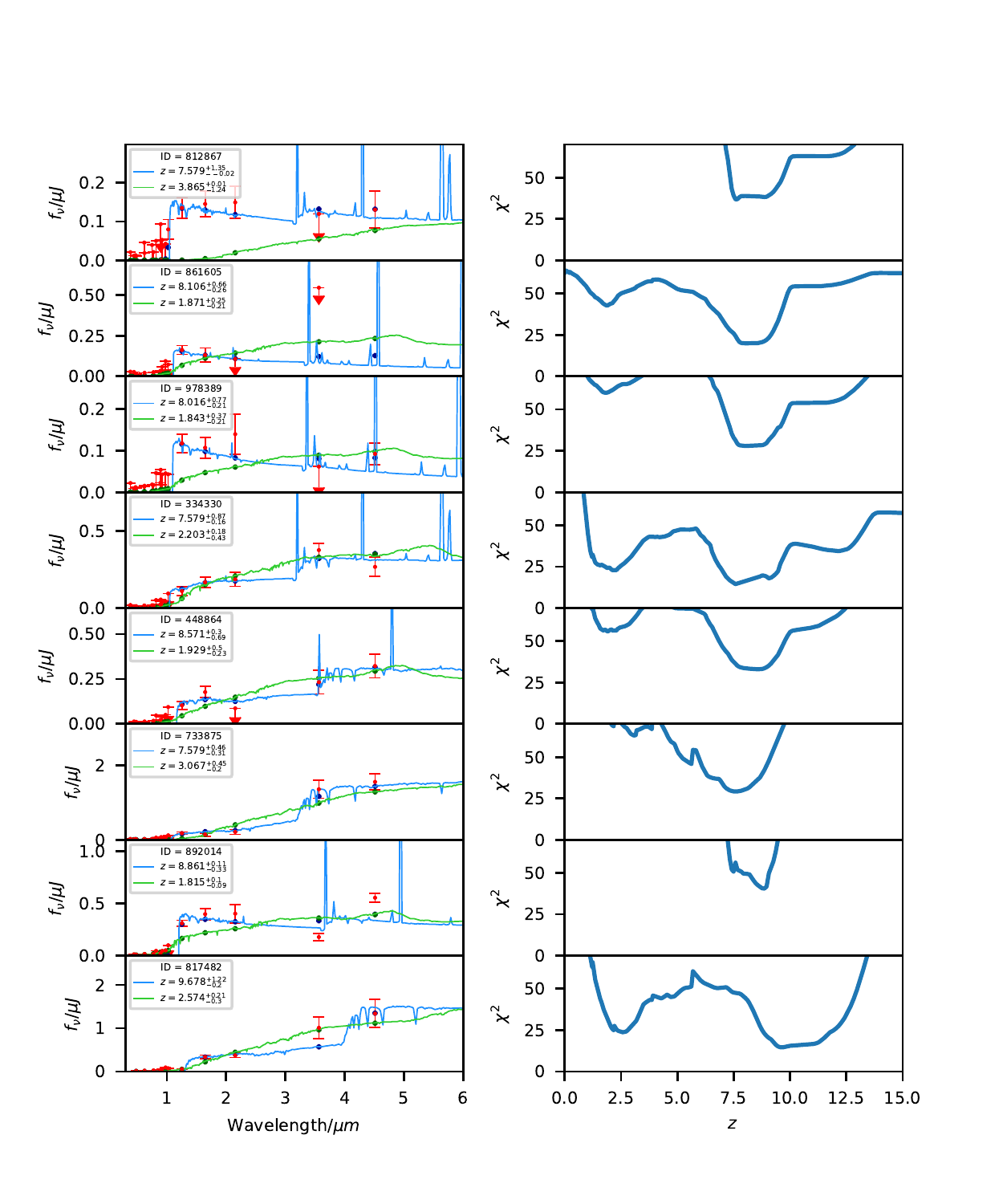}
    \caption{Continued.}
    \label{fig:COSMOS_SEDs_2}
\end{figure*}

\begin{figure*}
	\includegraphics[width=\textwidth]{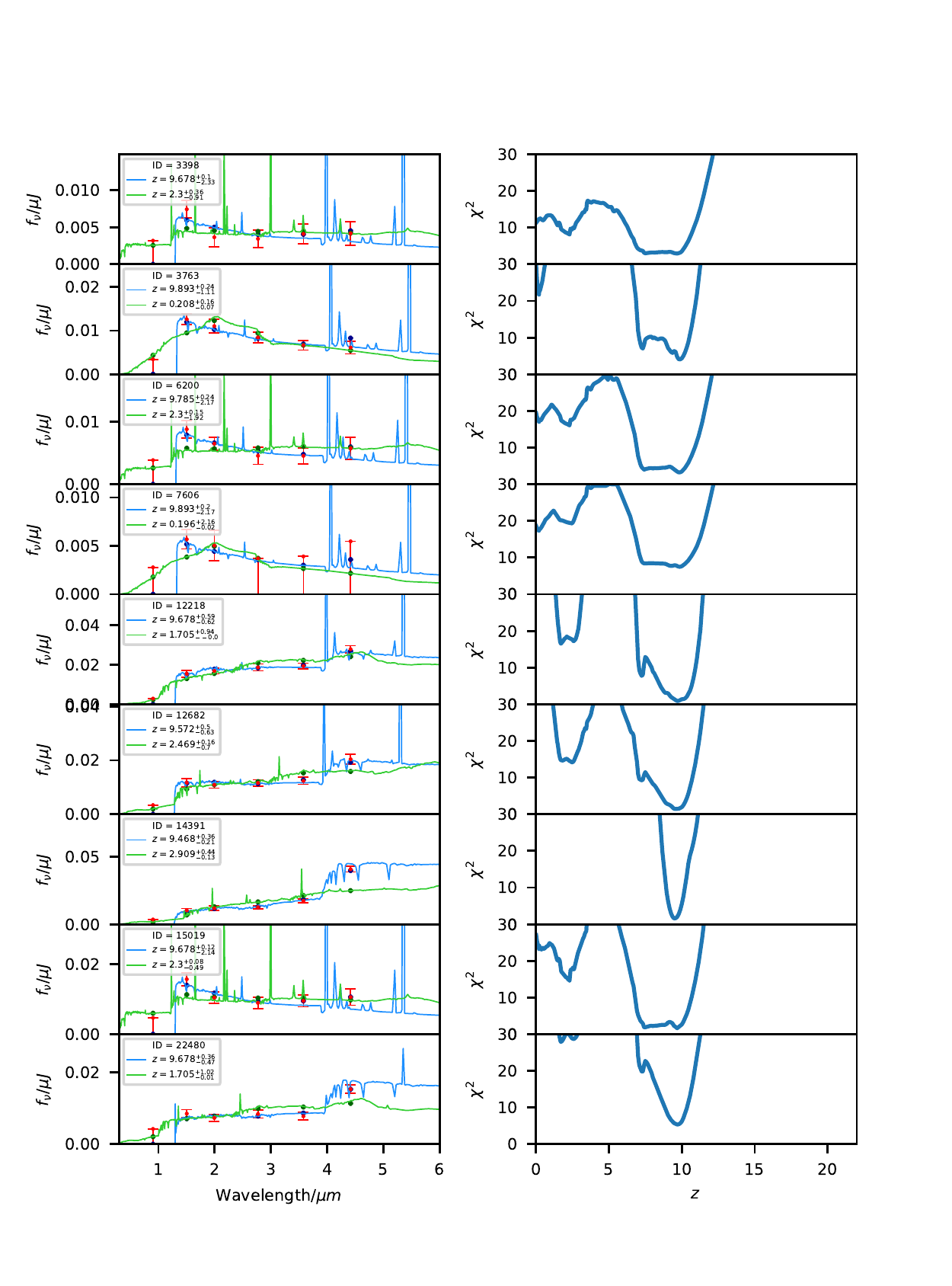}
    \caption{The measured photometry and best fitting SED from \texttt{EAZY} for the galaxies in the JWST sample. The format is the same as in Fig.\,\ref{fig:COSMOS_SEDs_1}.}
    \label{fig:jwst_SEDs1}
\end{figure*}

\begin{figure*}
	\includegraphics[width=\textwidth]{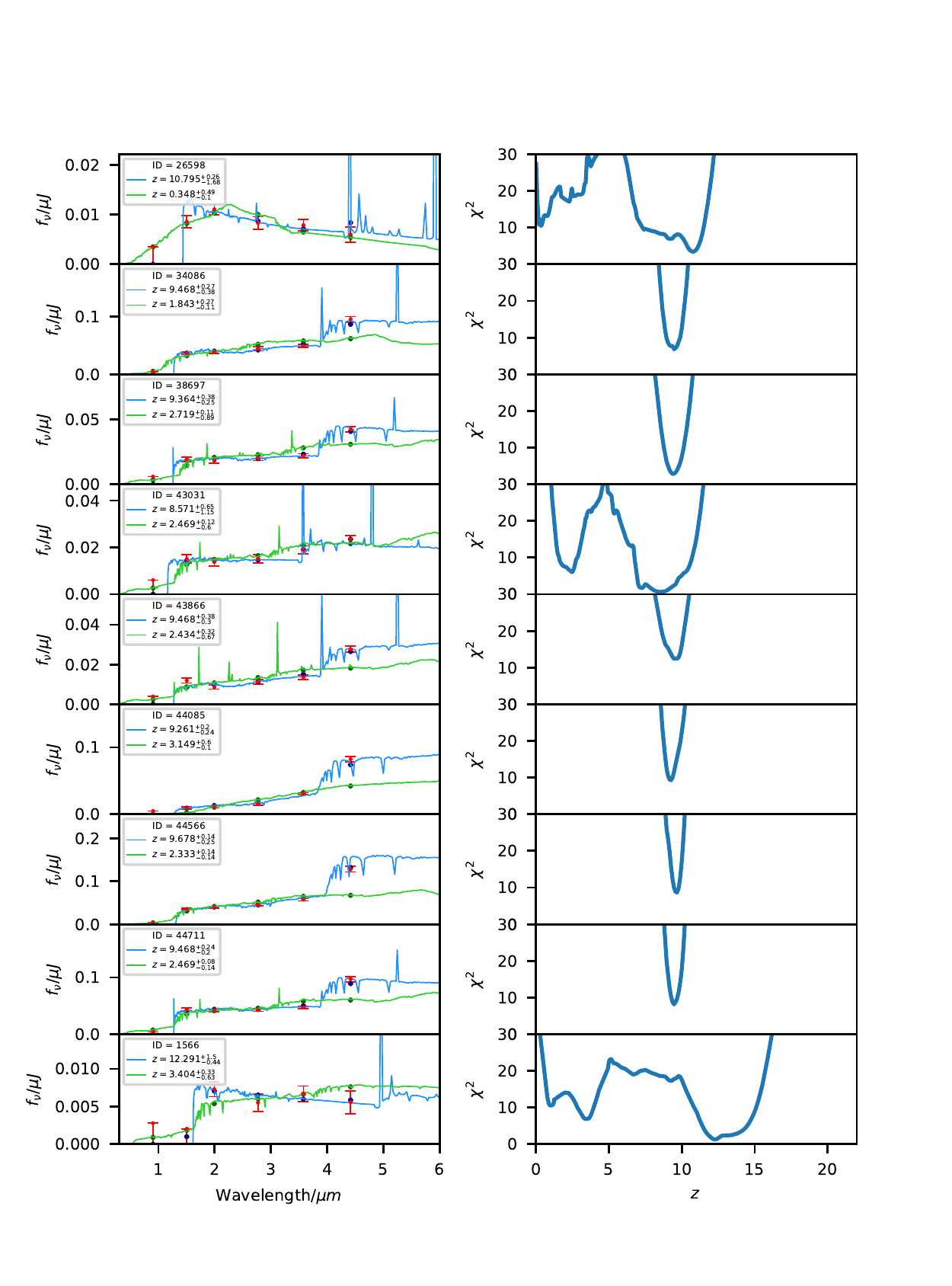}
    \caption{Continued.}
    \label{fig:jwst_SEDs2}
\end{figure*}

\begin{figure*}
	\includegraphics[width=\textwidth]{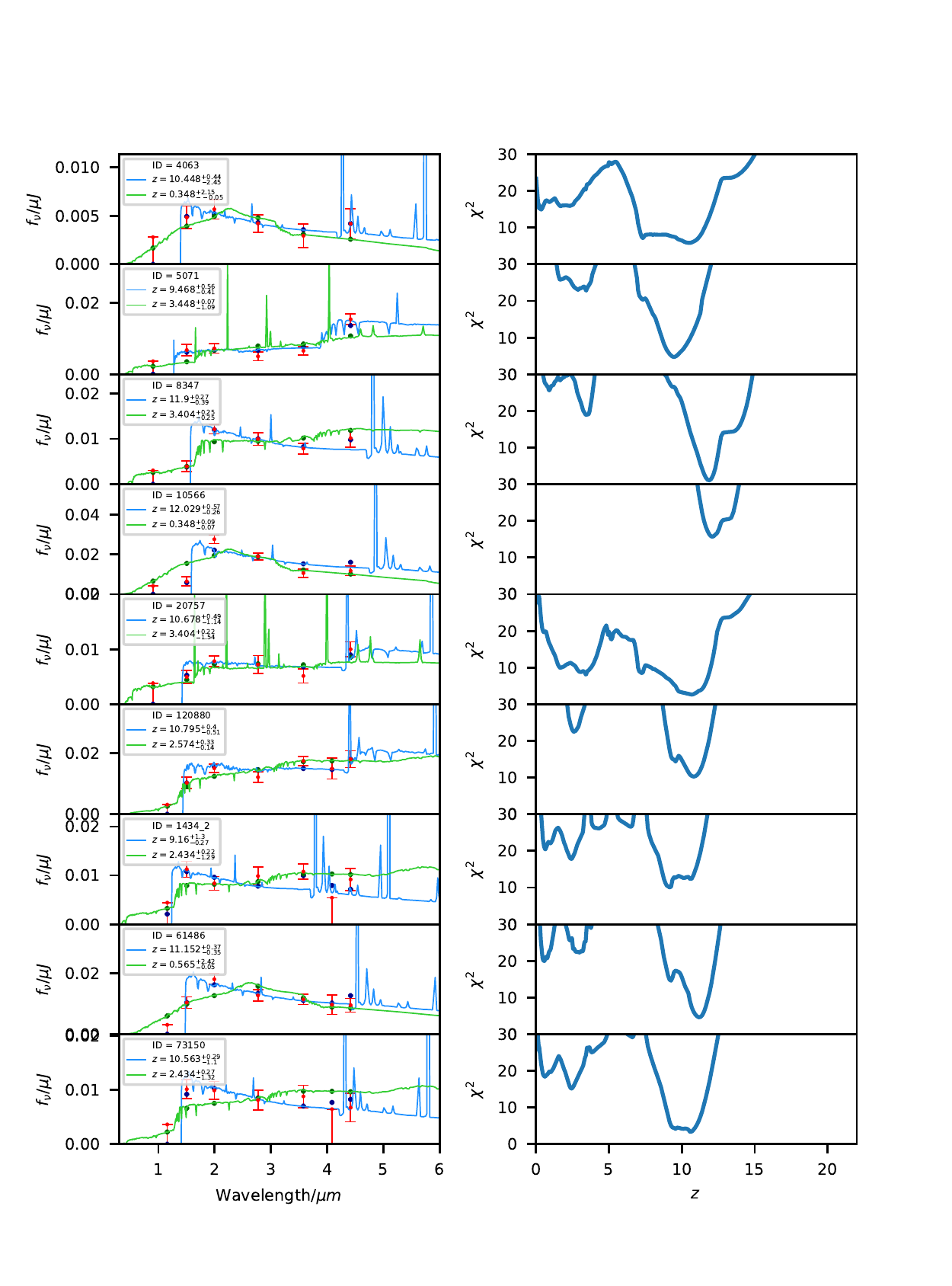}
    \caption{Continued.}
    \label{fig:jwst_SEDs3}
\end{figure*}

\begin{figure*}
	\includegraphics[width=\textwidth]{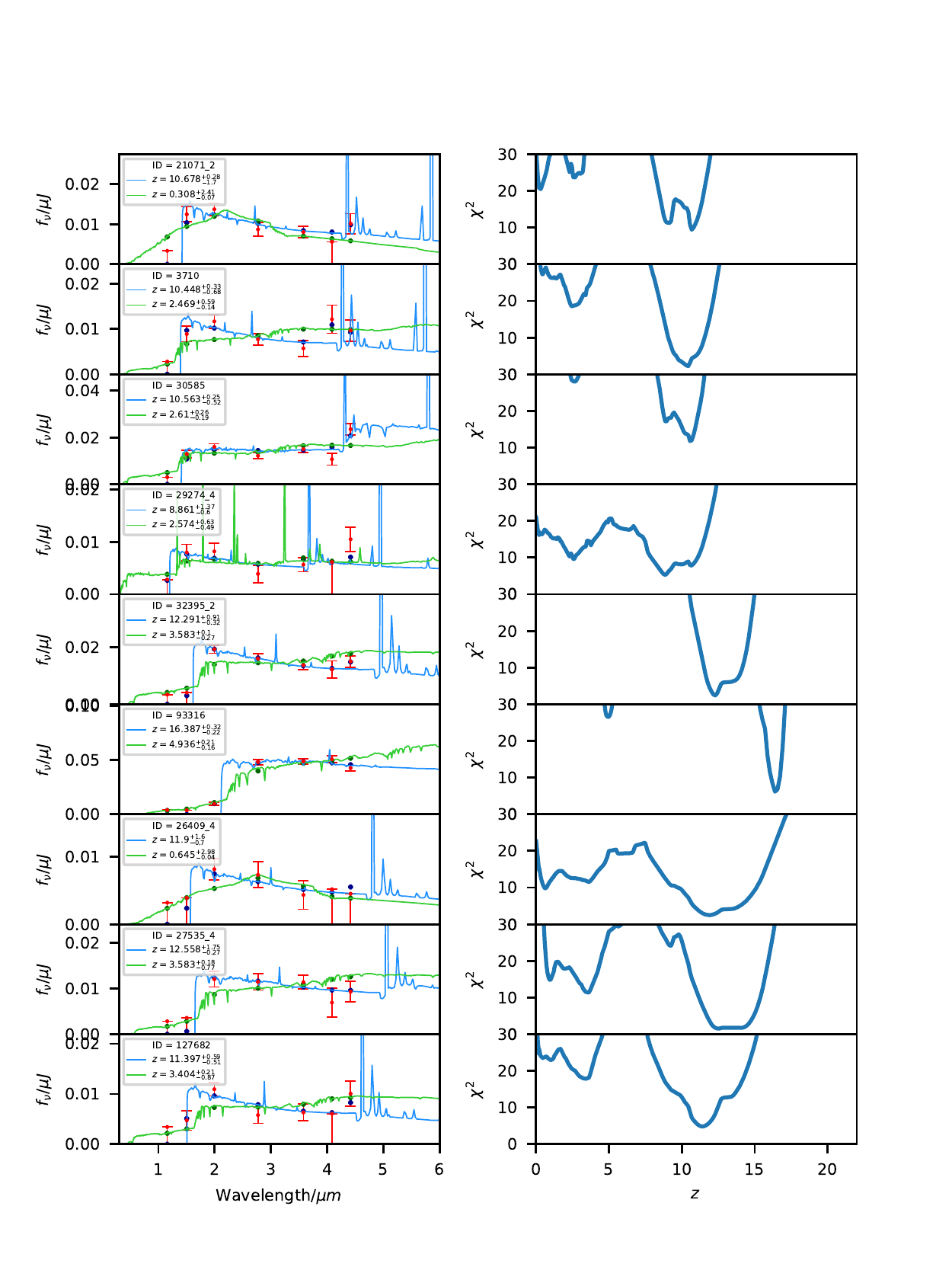}
    \caption{Continued.}
    \label{fig:jwst_SEDs4}
\end{figure*}

\begin{figure*}
	\includegraphics[width=\textwidth]{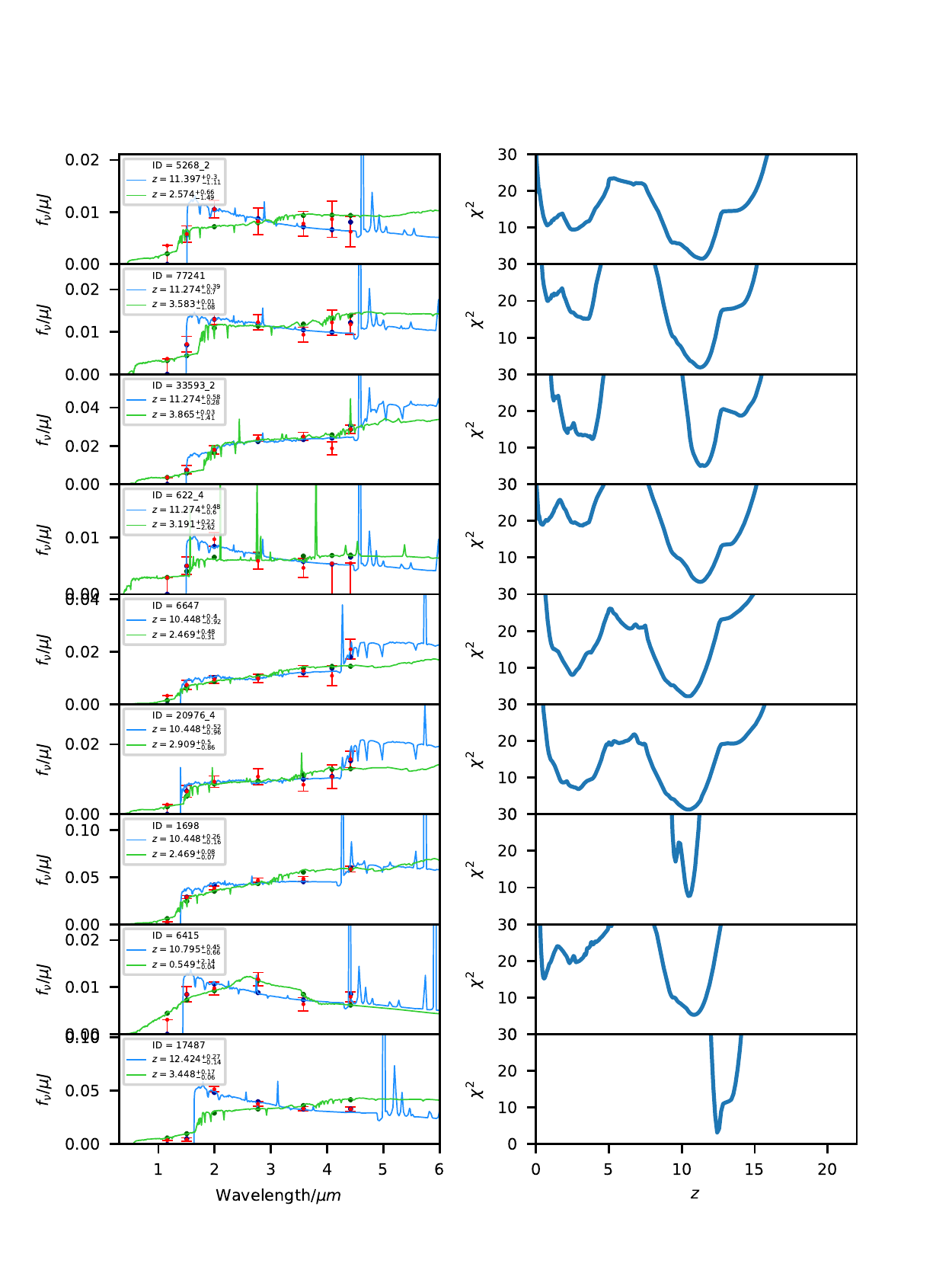}
    \caption{Continued.}
    \label{fig:jwst_SEDs5}
\end{figure*}

\begin{figure*}
	\includegraphics[width=\textwidth]{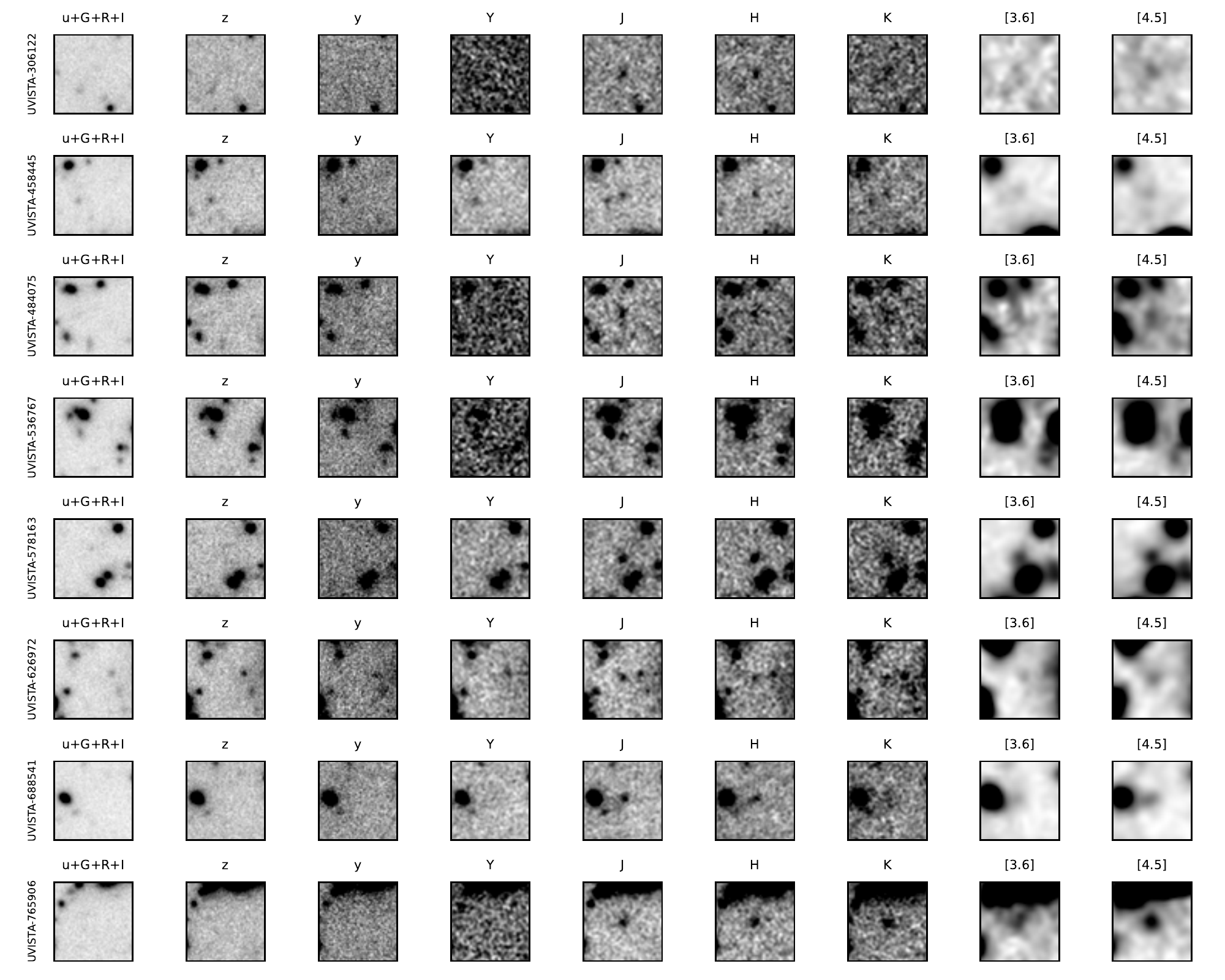}
    \caption{Postage-stamp images of the 16 $z>7.5$ galaxies selected from the COSMOS/UltraVISTA field. Each row shows an individual object, with the imaging ordered by increasing wavelength from left to right. Each postage-stamp image is 10 $\times$ 10\,arcsec.}
    \label{fig:COSMOS_cuts_1}
\end{figure*}

\begin{figure*}
	\includegraphics[width=\textwidth]{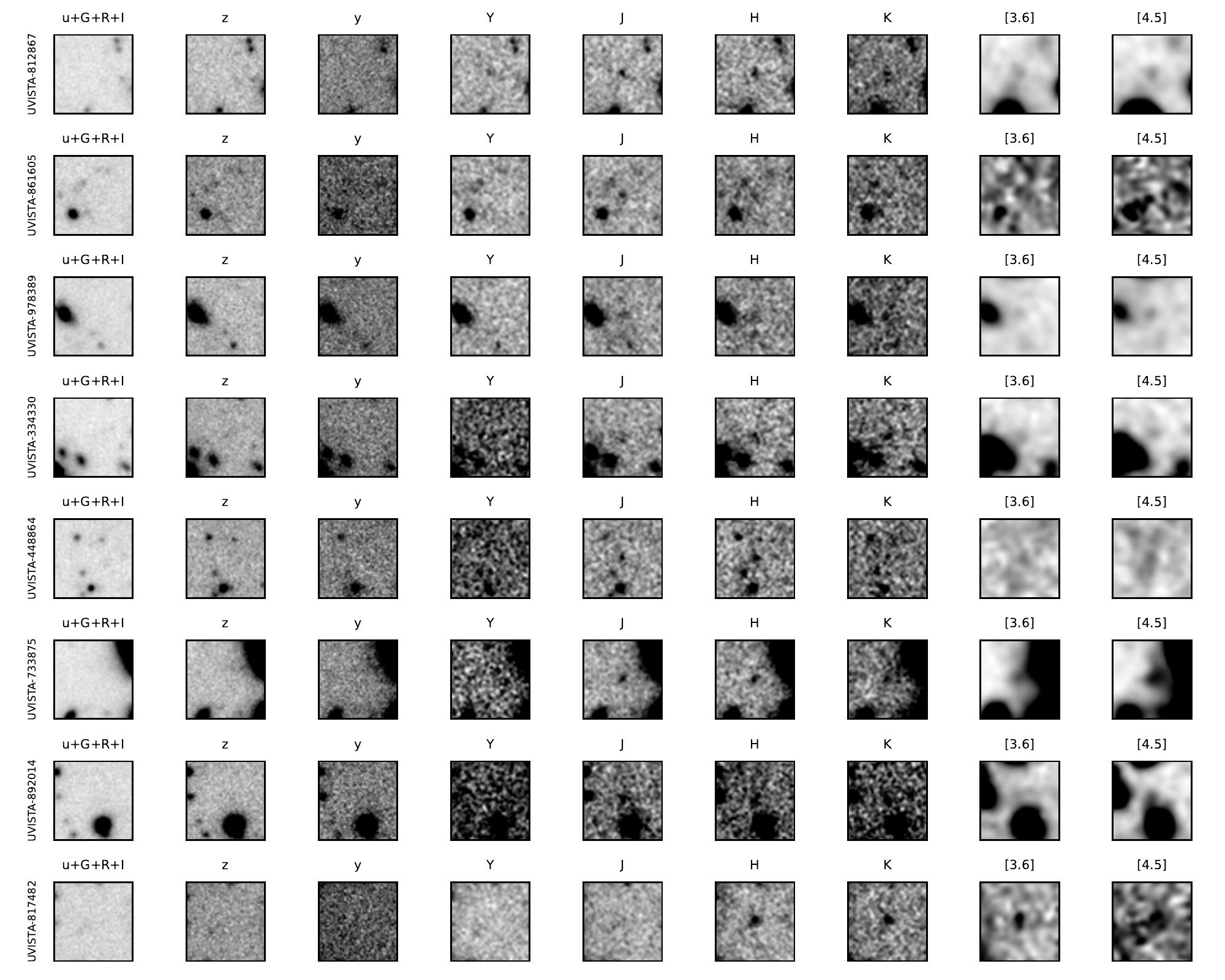}
    \caption{Continued.}
    \label{fig:COSMOS_cuts_2}
\end{figure*}

\begin{figure*}
	\includegraphics[width=\textwidth]{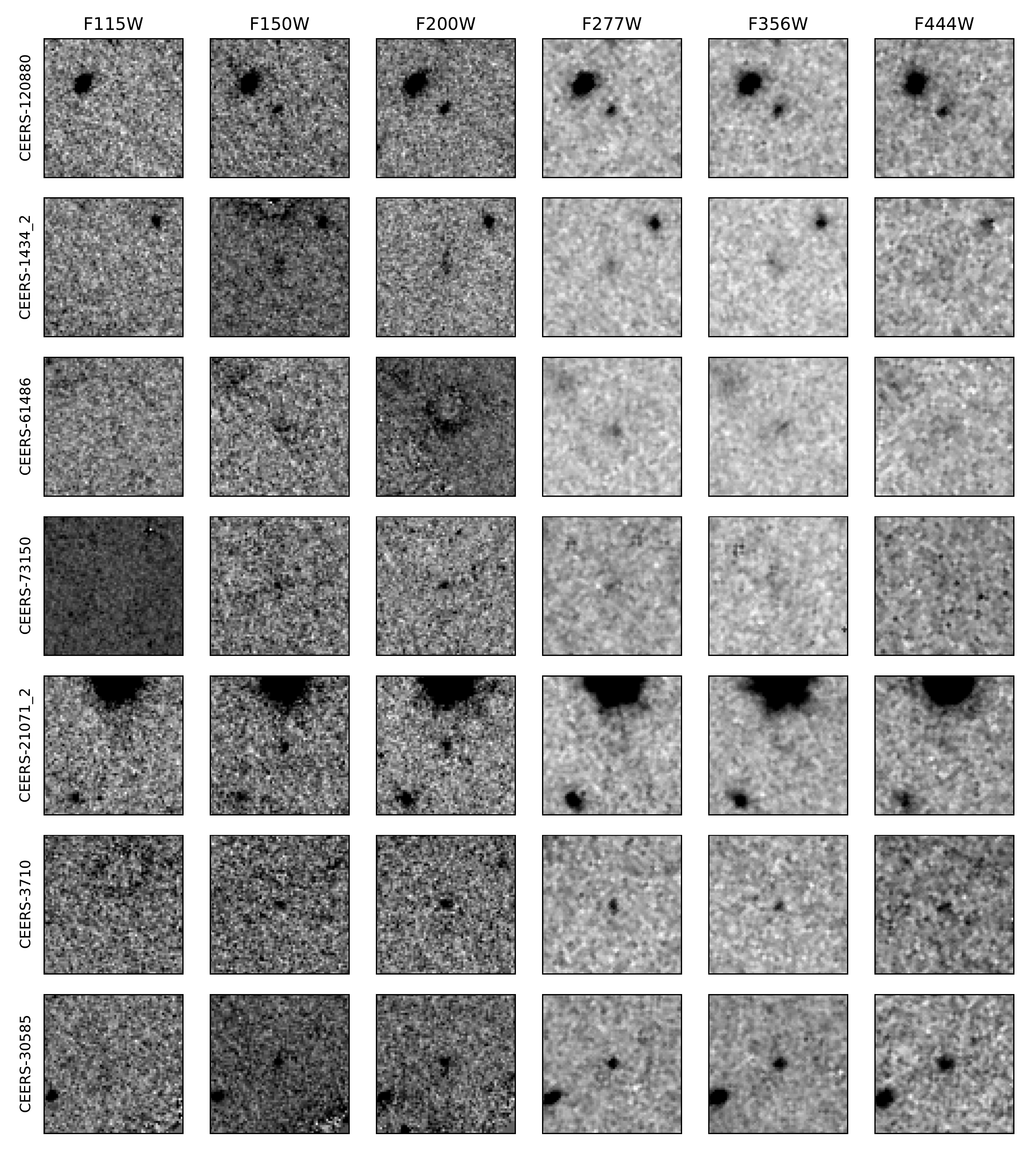}
    \caption{Postage-stamp images of the 45 $z>8.5$ galaxies selected from the combined JWST imaging. Each row shows an individual object, with the imaging ordered by increasing wavelength from left to right. Each postage-stamp image is 2 $\times$ 2\,arcsec.}
    \label{fig:jwst_cuts_1}
\end{figure*}

\begin{figure*}
	\includegraphics[width=\textwidth]{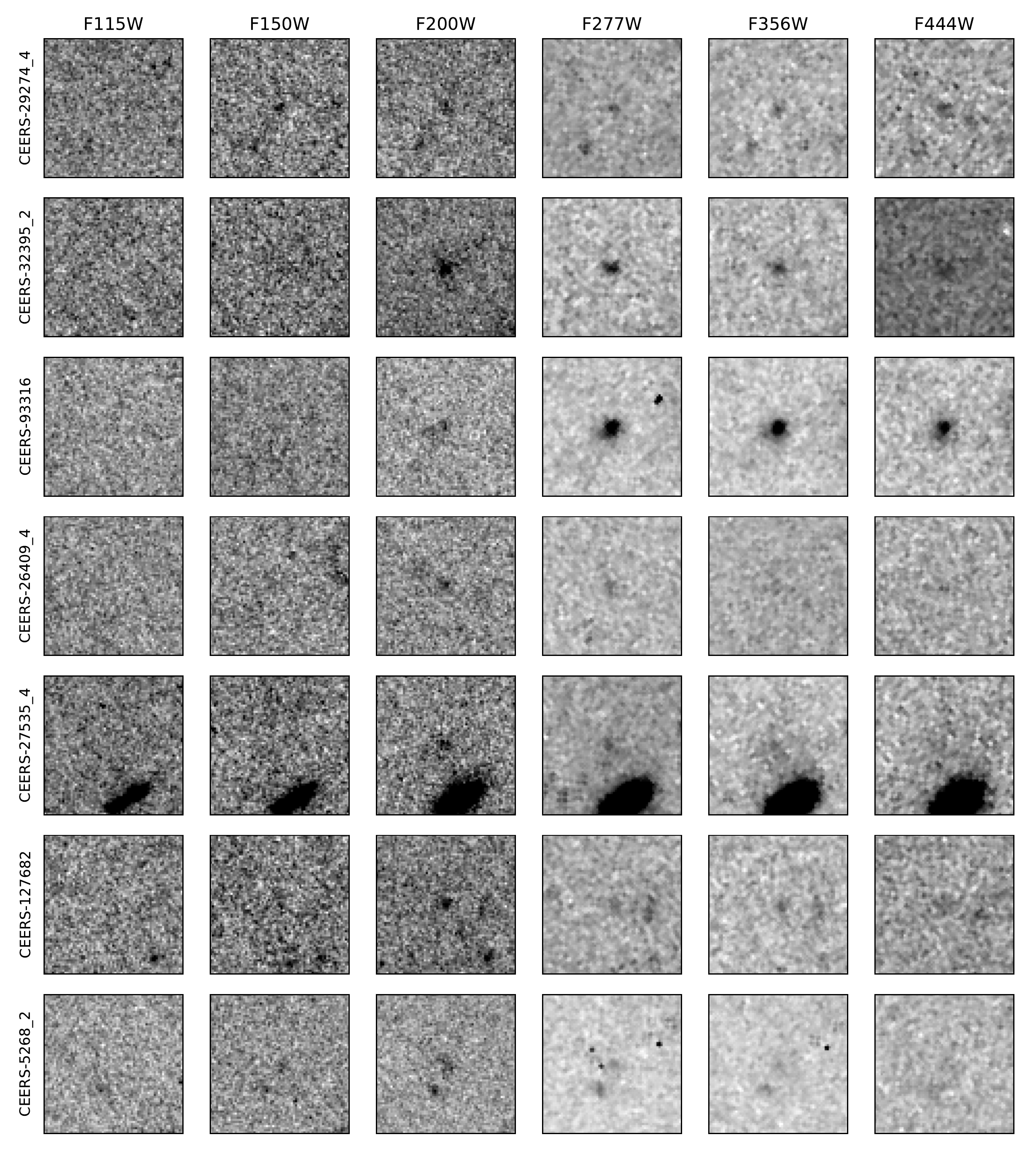}
    \caption{Continued.}
    \label{fig:jwst_cuts_2}
\end{figure*}

\begin{figure*}
	\includegraphics[width=\textwidth]{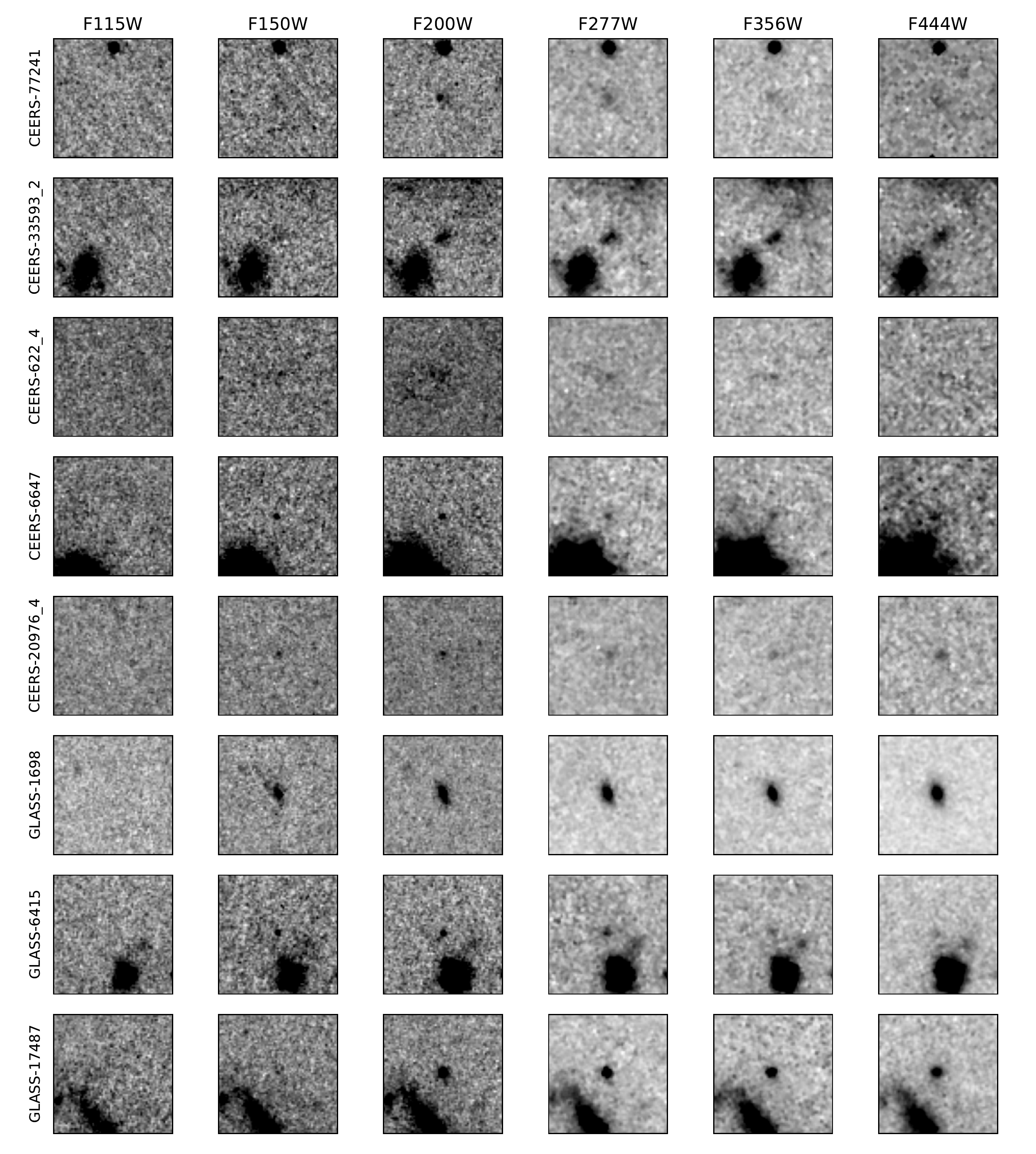}
    \caption{Continued.}
    \label{fig:jwst_cuts_3}
\end{figure*}

\begin{figure*}
	\includegraphics[width=\textwidth]{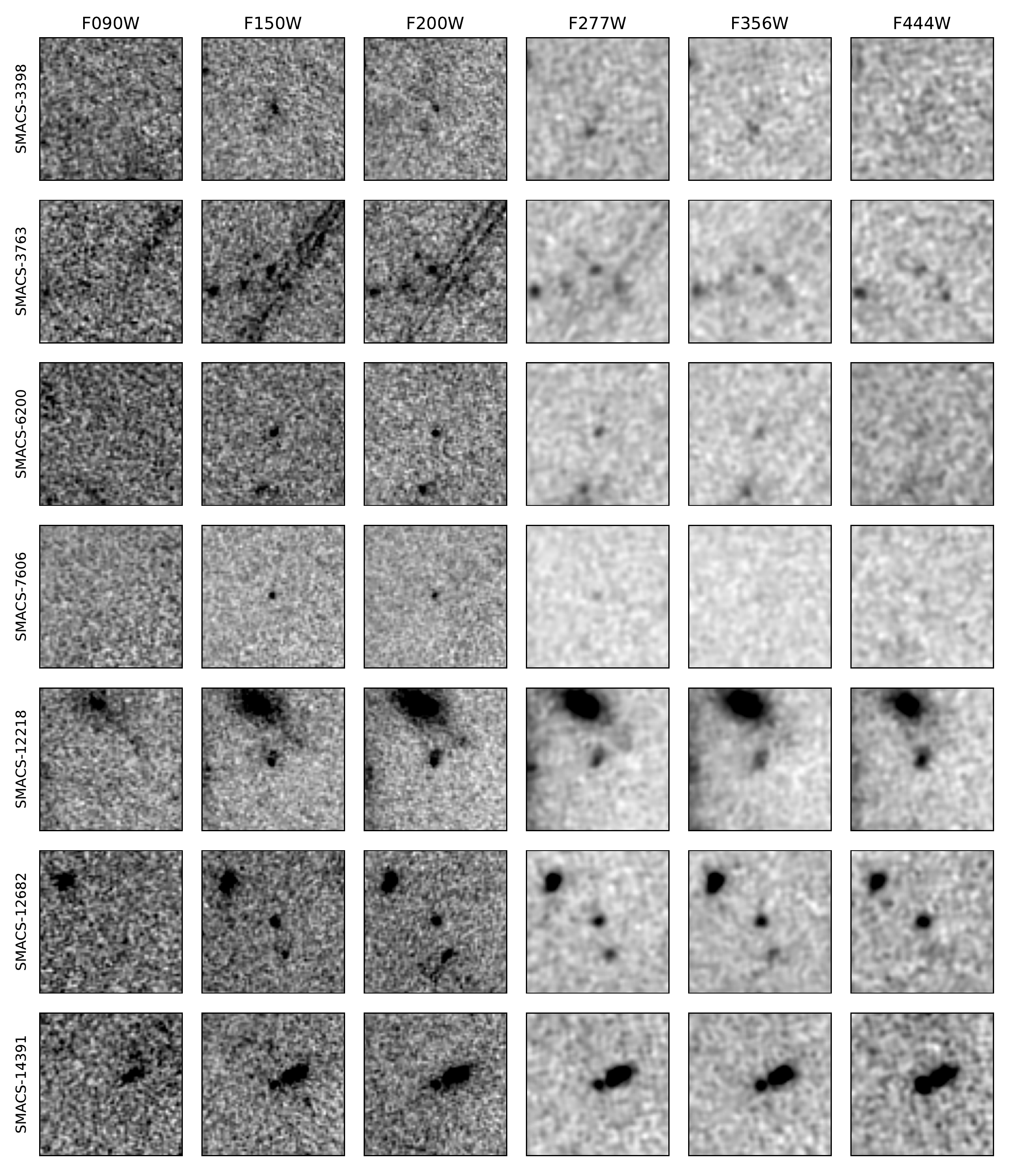}
    \caption{Continued.}
    \label{fig:jwst_cuts_4}
\end{figure*}

\begin{figure*}
	\includegraphics[width=\textwidth]{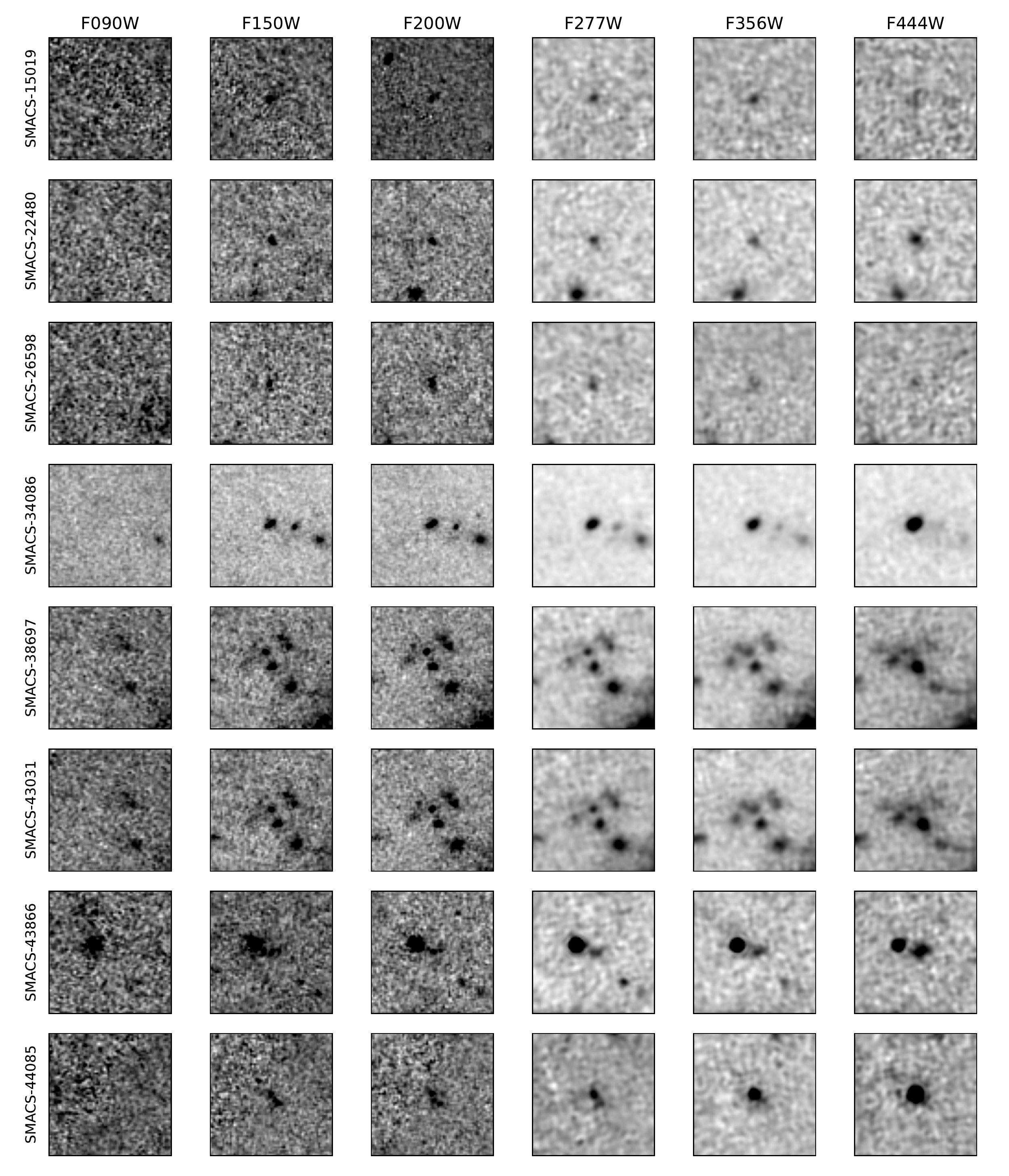}
    \caption{Continued.}
    \label{fig:jwst_cuts_5}
\end{figure*}

\begin{figure*}
	\includegraphics[width=\textwidth]{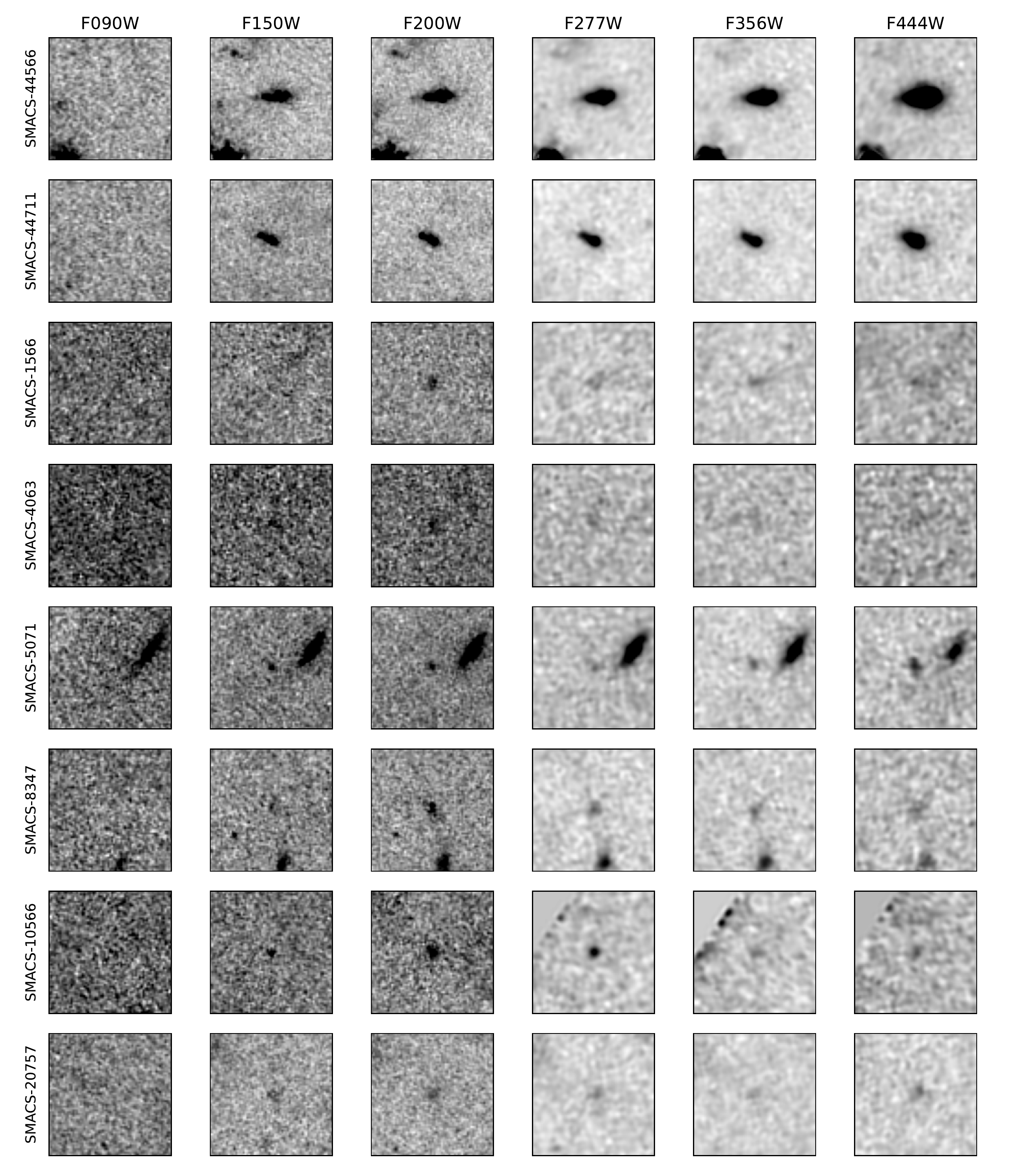}
    \caption{Continued.}
    \label{fig:jwst_cuts_6}
\end{figure*}

\section{Photometry tables}
The multi-wavelength photometry for the galaxies in the COSMOS/UltraVISTA sample is listed in Table \ref{tab:cosmos_phot}, with the photometry for the JWST-selected galaxies listed in Table \ref{tab:jwst_phot}.

\begin{table*}
	\centering
	\caption{The observed photometry for the galaxies selected from the COSMOS UltraVISTA DR5 imaging. The first column lists the ID of the object followed by the field within which it was identified. The third column shows the HSC $z-$band magnitude. The following columns show the HSC $y$ magnitude followed by VISTA $Y$ and VISTA $J, H, K_s$. The final columns show the \textit{Spitzer}/IRAC magnitudes. In the case of a non-detection, the 2$\sigma$ upper limit to the photometry is given.}
	\def\arraystretch{1.5}%
	\begin{tabular}{ c | c | c | c | c | c | c | c | c | c}
		\hline
		ID & FIELD & z & y & Y & J & H & K & [3.6] & [4.5] \\
		\hline
		306122 & COSMOS & >27.45 & >26.79 & >25.85 & 25.39$^{+0.14}_{-0.13}$ & 25.35$^{+0.23}_{-0.19}$ & 25.30$^{+0.29}_{-0.23}$ & >26.09 & 24.58$^{+0.1}_{-0.09}$ \\
		458445 & COSMOS & >27.33 & >26.60 & >26.70 & 25.63$^{+0.22}_{-0.18}$ & 26.01$^{+0.44}_{-0.31}$ & 25.90$^{+0.54}_{-0.36}$ & 25.27$^{+0.26}_{-0.21}$ & 24.81$^{+0.13}_{-0.12}$ \\
		484075 & COSMOS & >27.42 & >26.53 & >25.54 & 24.91$^{+0.14}_{-0.13}$ & 25.19$^{+0.23}_{-0.19}$ & 24.87$^{+0.26}_{-0.21}$ & 24.24$^{+0.24}_{-0.19}$ & 23.58$^{+0.13}_{-0.12}$ \\
		536767 & COSMOS & >27.34 & >26.61 & >25.75 & 25.44$^{+0.24}_{-0.2}$ & 25.77$^{+0.38}_{-0.28}$ & 25.34$^{+0.39}_{-0.29}$ & 23.77$^{+0.18}_{-0.15}$ & 23.29$^{+0.08}_{-0.07}$ \\
		578163 & COSMOS & >27.49 & >26.66 & >26.58 & 25.03$^{+0.14}_{-0.12}$ & 24.69$^{+0.12}_{-0.1}$ & 24.46$^{+0.15}_{-0.13}$ & 23.97$^{+0.08}_{-0.07}$ & 23.72$^{+0.06}_{-0.05}$ \\
		626972 & COSMOS & >27.33 & >26.52 & >26.47 & 25.42$^{+0.21}_{-0.17}$ & 25.67$^{+0.4}_{-0.29}$ & 25.68$^{+0.57}_{-0.37}$ & 24.30$^{+0.42}_{-0.3}$ & 24.03$^{+0.24}_{-0.19}$ \\
		688541 & COSMOS & >27.14 & >26.55 & >26.65 & 24.61$^{+0.09}_{-0.08}$ & 24.98$^{+0.15}_{-0.14}$ & 24.96$^{+0.27}_{-0.22}$ & 24.47$^{+0.19}_{-0.16}$ & 23.89$^{+0.11}_{-0.1}$ \\
		765906 & COSMOS & >27.21 & >26.82 & >26.04 & 24.57$^{+0.1}_{-0.09}$ & 24.46$^{+0.1}_{-0.09}$ & 24.11$^{+0.11}_{-0.1}$ & 23.07$^{+0.17}_{-0.15}$ & 22.95$^{+0.08}_{-0.07}$ \\
		812867 & COSMOS & >27.44 & >26.87 & 26.24$^{+0.42}_{-0.3}$ & 25.70$^{+0.24}_{-0.2}$ & 25.62$^{+0.28}_{-0.22}$ & 25.60$^{+0.35}_{-0.27}$ & >25.84 & 25.74$^{+0.49}_{-0.34}$ \\
		861605 & COSMOS & >27.14 & >26.02 & >26.33 & 25.50$^{+0.2}_{-0.17}$ & 25.74$^{+0.45}_{-0.32}$ & >25.96 & >24.18 & >23.71 \\
		978389 & COSMOS & >27.59 & >26.81 & >26.91 & 25.86$^{+0.23}_{-0.19}$ & 25.96$^{+0.29}_{-0.23}$ & 25.67$^{+0.46}_{-0.32}$ & >26.56 & 26.12$^{+0.36}_{-0.27}$ \\
		334330 & COSMOS & >27.54 & >26.85 & >26.05 & 25.92$^{+0.31}_{-0.24}$ & 25.47$^{+0.24}_{-0.19}$ & 25.35$^{+0.31}_{-0.24}$ & 24.59$^{+0.14}_{-0.12}$ & 24.96$^{+0.29}_{-0.23}$ \\
		448864 & COSMOS & >27.57 & >26.69 & >26.05 & 26.00$^{+0.27}_{-0.22}$ & 25.41$^{+0.21}_{-0.18}$ & >26.18 & 25.11$^{+0.36}_{-0.27}$ & 24.76$^{+0.25}_{-0.2}$ \\
		733875 & COSMOS & >27.33 & >26.43 & >25.78 & 25.39$^{+0.2}_{-0.17}$ & 25.64$^{+0.39}_{-0.29}$ & 25.22$^{+0.4}_{-0.29}$ & 23.19$^{+0.21}_{-0.18}$ & 23.04$^{+0.16}_{-0.14}$ \\
		892014 & COSMOS & >27.63 & >26.68 & >26.00 & 24.80$^{+0.11}_{-0.1}$ & 24.53$^{+0.16}_{-0.14}$ & 24.52$^{+0.27}_{-0.21}$ & 25.40$^{+0.23}_{-0.19}$ & 24.17$^{+0.09}_{-0.08}$ \\
		817482 & COSMOS & >27.21 & >26.01 & >26.30 & >26.65 & 24.71$^{+0.15}_{-0.13}$ & 24.60$^{+0.15}_{-0.13}$ & 23.52$^{+0.31}_{-0.24}$ & 23.21$^{+0.3}_{-0.24}$ \\
	\end{tabular}
	\label{tab:cosmos_phot}
\end{table*}

\begin{table*}
	\centering
	\caption{The observed photometry for the galaxies in the JWST-selected sample. The table first presents the photometry for the SMACS0723 candidates followed by the CEERS objects, and then finally the 3 GLASS candidates. The first column lists the ID of the object followed by the name of the field in which it was  identified. The following columns show the photometry in each of the relevant NIRCam filters. A dash indicates where that filter was not available for a given field. In the case of a non-detection at the 2$\sigma$ level the photometry is shown as an upper limit. Extended sources are indicated with an asterisk beside the ID number. These sources have had an extra correction applied to their photometry as a point-source correction is insufficient. }
	\def\arraystretch{1.25}%
	\begin{tabular}{c | c | c | c | c | c | c | c | c | c}
		\hline
		ID & FIELD & F090W & F115W & F150W & F200W & F277W & F356W & F410M & F444W \\
		\hline
		3398 & SMACS & >29.85 & - & 28.93$^{+0.19}_{-0.16}$ & 29.70$^{+0.47}_{-0.33}$ & 29.76$^{+0.47}_{-0.33}$ & 29.57$^{+0.42}_{-0.3}$ & - & 29.55$^{+0.53}_{-0.35}$ \\
		3763 & SMACS & >29.77 & - & 28.35$^{+0.11}_{-0.1}$ & 28.49$^{+0.16}_{-0.14}$ & 28.79$^{+0.18}_{-0.15}$ & 29.04$^{+0.2}_{-0.17}$ & - & 29.13$^{+0.3}_{-0.23}$ \\
		6200 & SMACS & >29.64 & - & 28.75$^{+0.18}_{-0.16}$ & 29.08$^{+0.19}_{-0.16}$ & 29.46$^{+0.39}_{-0.29}$ & 29.46$^{+0.37}_{-0.27}$ & - & 29.21$^{+0.4}_{-0.29}$ \\
		7606 & SMACS & >29.99 & - & 29.22$^{+0.21}_{-0.17}$ & 29.35$^{+0.4}_{-0.29}$ & >29.68 & >29.62 & - & >29.26 \\
		12218 & SMACS & >29.97 & - & 28.14$^{+0.12}_{-0.11}$ & 28.04$^{+0.1}_{-0.09}$ & 27.93$^{+0.11}_{-0.1}$ & 27.89$^{+0.08}_{-0.08}$ & - & 27.50$^{+0.07}_{-0.07}$ \\
		12682 & SMACS & >29.78 & - & 28.43$^{+0.15}_{-0.13}$ & 28.51$^{+0.13}_{-0.12}$ & 28.44$^{+0.12}_{-0.11}$ & 28.37$^{+0.12}_{-0.11}$ & - & 27.83$^{+0.1}_{-0.09}$ \\
		14391 & SMACS & >29.73 & - & 28.56$^{+0.15}_{-0.13}$ & 28.42$^{+0.16}_{-0.14}$ & 28.34$^{+0.1}_{-0.09}$ & 28.00$^{+0.08}_{-0.08}$ & - & 27.07$^{+0.06}_{-0.05}$ \\
		15019 & SMACS & >29.43 & - & 28.11$^{+0.14}_{-0.12}$ & 28.55$^{+0.19}_{-0.16}$ & 28.72$^{+0.22}_{-0.18}$ & 28.65$^{+0.21}_{-0.18}$ & - & 28.54$^{+0.27}_{-0.21}$ \\
		22480 & SMACS & >29.53 & - & 28.78$^{+0.16}_{-0.14}$ & 28.95$^{+0.15}_{-0.14}$ & 28.79$^{+0.16}_{-0.14}$ & 28.88$^{+0.16}_{-0.14}$ & - & 28.14$^{+0.08}_{-0.08}$ \\
		26598 & SMACS & >29.74 & - & 28.77$^{+0.17}_{-0.15}$ & 28.49$^{+0.11}_{-0.1}$ & 28.81$^{+0.18}_{-0.15}$ & 28.86$^{+0.18}_{-0.15}$ & - & 29.16$^{+0.33}_{-0.25}$ \\
		34086 & SMACS & >29.62 & - & 27.19$^{+0.06}_{-0.06}$ & 27.15$^{+0.06}_{-0.06}$ & 27.00$^{+0.09}_{-0.08}$ & 26.88$^{+0.06}_{-0.06}$ & - & 26.16$^{+0.06}_{-0.05}$ \\
		38697 & SMACS & >29.17 & - & 27.90$^{+0.11}_{-0.1}$ & 27.96$^{+0.11}_{-0.1}$ & 27.85$^{+0.1}_{-0.09}$ & 27.75$^{+0.09}_{-0.08}$ & - & 27.03$^{+0.06}_{-0.05}$ \\
		43031 & SMACS & >29.15 & - & 28.15$^{+0.14}_{-0.12}$ & 28.26$^{+0.15}_{-0.13}$ & 28.15$^{+0.13}_{-0.12}$ & 27.91$^{+0.1}_{-0.09}$ & - & 27.67$^{+0.07}_{-0.07}$ \\
		43866 & SMACS & >29.61 & - & 28.41$^{+0.12}_{-0.11}$ & 28.76$^{+0.14}_{-0.12}$ & 28.48$^{+0.13}_{-0.11}$ & 28.28$^{+0.1}_{-0.09}$ & - & 27.50$^{+0.06}_{-0.06}$ \\
		44085 & SMACS & >29.39 & - & 28.70$^{+0.22}_{-0.18}$ & 28.53$^{+0.14}_{-0.12}$ & 28.17$^{+0.1}_{-0.09}$ & 27.38$^{+0.06}_{-0.05}$ & - & 26.31$^{+0.06}_{-0.05}$ \\
		44566$^*$ & SMACS & >29.84 & - & 26.24$^{+0.06}_{-0.05}$ & 26.11$^{+0.06}_{-0.05}$ & 25.98$^{+0.06}_{-0.05}$ & 25.72$^{+0.06}_{-0.05}$ & - & 24.84$^{+0.06}_{-0.05}$ \\
		44711$^*$ & SMACS & >29.58 & - & 26.63$^{+0.06}_{-0.05}$ & 26.68$^{+0.06}_{-0.05}$ & 26.66$^{+0.06}_{-0.05}$ & 26.55$^{+0.06}_{-0.05}$ & - & 25.77$^{+0.06}_{-0.05}$ \\
		1566 & SMACS & >29.99 & - & >30.33 & 28.94$^{+0.16}_{-0.14}$ & 29.25$^{+0.25}_{-0.21}$ & 29.04$^{+0.18}_{-0.16}$ & - & 29.24$^{+0.34}_{-0.26}$ \\
		4063 & SMACS & >29.99 & - & 29.39$^{+0.3}_{-0.23}$ & 29.22$^{+0.21}_{-0.18}$ & 29.54$^{+0.26}_{-0.21}$ & 29.93$^{+0.56}_{-0.37}$ & - & 29.55$^{+0.51}_{-0.35}$ \\
		5071 & SMACS & >29.70 & - & 29.02$^{+0.28}_{-0.22}$ & 28.95$^{+0.22}_{-0.18}$ & 29.35$^{+0.31}_{-0.24}$ & 29.06$^{+0.2}_{-0.17}$ & - & 28.13$^{+0.11}_{-0.1}$ \\
		8347 & SMACS & >29.89 & - & 29.61$^{+0.38}_{-0.28}$ & 28.39$^{+0.1}_{-0.09}$ & 28.60$^{+0.16}_{-0.14}$ & 28.86$^{+0.18}_{-0.15}$ & - & 28.59$^{+0.23}_{-0.19}$ \\
		10566 & SMACS & >29.56 & - & 29.08$^{+0.47}_{-0.33}$ & 27.50$^{+0.09}_{-0.08}$ & 27.90$^{+0.1}_{-0.1}$ & 28.54$^{+0.23}_{-0.19}$ & - & 28.42$^{+0.24}_{-0.2}$ \\
		20757 & SMACS & >29.64 & - & 29.36$^{+0.29}_{-0.23}$ & 28.87$^{+0.15}_{-0.13}$ & 28.95$^{+0.28}_{-0.22}$ & 29.32$^{+0.31}_{-0.24}$ & - & 28.60$^{+0.16}_{-0.14}$ \\
		\hline
		120880 & CEERS & - & >29.87 & 28.58$^{+0.23}_{-0.19}$ & 28.16$^{+0.11}_{-0.1}$ & 28.40$^{+0.16}_{-0.14}$ & 28.00$^{+0.1}_{-0.09}$ & 28.17$^{+0.28}_{-0.22}$ & 27.97$^{+0.18}_{-0.15}$ \\
		1434\_2 & CEERS & - & >29.49 & 28.47$^{+0.17}_{-0.15}$ & 28.80$^{+0.2}_{-0.17}$ & 28.62$^{+0.22}_{-0.18}$ & 28.52$^{+0.17}_{-0.14}$ & >29.27 & 28.70$^{+0.31}_{-0.24}$ \\
		61486 & CEERS & - & >29.88 & 28.56$^{+0.22}_{-0.18}$ & 27.96$^{+0.11}_{-0.1}$ & 28.34$^{+0.18}_{-0.15}$ & 28.46$^{+0.17}_{-0.15}$ & 28.63$^{+0.44}_{-0.31}$ & 28.65$^{+0.29}_{-0.23}$ \\
		73150 & CEERS & - & >29.70 & 28.58$^{+0.21}_{-0.17}$ & 28.61$^{+0.2}_{-0.17}$ & 28.82$^{+0.28}_{-0.22}$ & 28.74$^{+0.29}_{-0.23}$ & >29.08 & 29.03$^{+0.53}_{-0.36}$ \\
		21071\_2 & CEERS & - & >29.79 & 28.36$^{+0.18}_{-0.15}$ & 28.26$^{+0.13}_{-0.12}$ & 28.75$^{+0.24}_{-0.19}$ & 28.84$^{+0.21}_{-0.18}$ & >29.23 & 28.59$^{+0.31}_{-0.24}$ \\
		3710 & CEERS & - & >29.99 & 28.73$^{+0.23}_{-0.19}$ & 28.43$^{+0.15}_{-0.13}$ & 28.88$^{+0.2}_{-0.17}$ & 29.21$^{+0.4}_{-0.29}$ & 28.39$^{+0.32}_{-0.25}$ & 28.64$^{+0.3}_{-0.24}$ \\
		30585 & CEERS & - & >29.91 & 28.32$^{+0.14}_{-0.13}$ & 28.09$^{+0.09}_{-0.08}$ & 28.39$^{+0.12}_{-0.11}$ & 28.16$^{+0.12}_{-0.11}$ & 28.53$^{+0.29}_{-0.23}$ & 27.67$^{+0.12}_{-0.11}$ \\
		29274\_4 & CEERS & - & >29.98 & 28.84$^{+0.24}_{-0.19}$ & 28.82$^{+0.23}_{-0.19}$ & 29.63$^{+0.61}_{-0.39}$ & 29.22$^{+0.3}_{-0.24}$ & >29.13 & 28.55$^{+0.27}_{-0.22}$ \\
		32395\_2 & CEERS & - & >29.78 & >29.58 & 27.89$^{+0.08}_{-0.08}$ & 28.09$^{+0.12}_{-0.11}$ & 28.26$^{+0.13}_{-0.12}$ & 28.38$^{+0.3}_{-0.24}$ & 28.16$^{+0.16}_{-0.14}$ \\
		93316$^*$ & CEERS & - & >29.63 & >29.68 & 28.25$^{+0.18}_{-0.16}$ & 26.49$^{+0.06}_{-0.05}$ & 26.48$^{+0.06}_{-0.05}$ & 26.44$^{+0.07}_{-0.07}$ & 26.63$^{+0.07}_{-0.06}$ \\
		26409\_4 & CEERS & - & >29.84 & >29.62 & 28.82$^{+0.23}_{-0.19}$ & 28.94$^{+0.33}_{-0.25}$ & 29.51$^{+0.73}_{-0.43}$ & >29.31 & >29.46 \\
		27535\_4 & CEERS & - & >29.97 & >29.72 & 28.40$^{+0.16}_{-0.14}$ & 28.45$^{+0.18}_{-0.16}$ & 28.46$^{+0.15}_{-0.13}$ & 29.00$^{+0.65}_{-0.4}$ & 28.68$^{+0.3}_{-0.23}$ \\
		127682 & CEERS & - & >29.76 & 29.41$^{+0.58}_{-0.37}$ & 28.50$^{+0.14}_{-0.12}$ & 29.19$^{+0.36}_{-0.27}$ & 29.10$^{+0.34}_{-0.26}$ & >29.15 & 28.60$^{+0.31}_{-0.24}$ \\
		5268\_2 & CEERS & - & >29.72 & 29.19$^{+0.34}_{-0.26}$ & 28.54$^{+0.19}_{-0.16}$ & 28.81$^{+0.4}_{-0.29}$ & 28.87$^{+0.39}_{-0.29}$ & 28.76$^{+0.55}_{-0.37}$ & 29.11$^{+0.69}_{-0.42}$ \\
		77241$^*$ & CEERS & - & >29.69 & 28.70$^{+0.32}_{-0.25}$ & 28.05$^{+0.11}_{-0.1}$ & 28.11$^{+0.17}_{-0.15}$ & 28.41$^{+0.23}_{-0.19}$ & 28.12$^{+0.3}_{-0.23}$ & 28.15$^{+0.25}_{-0.2}$ \\
		33593\_2 & CEERS & - & >29.73 & 28.90$^{+0.35}_{-0.26}$ & 27.96$^{+0.14}_{-0.12}$ & 27.65$^{+0.08}_{-0.07}$ & 27.61$^{+0.1}_{-0.09}$ & 27.92$^{+0.22}_{-0.18}$ & 27.46$^{+0.09}_{-0.08}$ \\
		622\_4 & CEERS & - & >29.93 & 29.35$^{+0.4}_{-0.29}$ & 28.63$^{+0.14}_{-0.13}$ & 29.17$^{+0.31}_{-0.24}$ & 29.43$^{+0.48}_{-0.33}$ & >29.26 & >29.25 \\
		6647 & CEERS & - & >29.83 & 28.93$^{+0.29}_{-0.23}$ & 28.66$^{+0.19}_{-0.16}$ & 28.62$^{+0.19}_{-0.16}$ & 28.34$^{+0.19}_{-0.16}$ & 28.51$^{+0.46}_{-0.32}$ & 27.80$^{+0.22}_{-0.18}$ \\
		20976\_4 & CEERS & - & >30.00 & 29.07$^{+0.32}_{-0.24}$ & 28.68$^{+0.21}_{-0.17}$ & 28.52$^{+0.25}_{-0.2}$ & 28.79$^{+0.27}_{-0.22}$ & 28.53$^{+0.42}_{-0.3}$ & 28.11$^{+0.18}_{-0.15}$ \\
		\hline
		1698$^*$ & GLASS & - & >30.19 & 26.89$^{+0.06}_{-0.06}$ & 26.88$^{+0.06}_{-0.05}$ & 26.67$^{+0.06}_{-0.05}$ & 26.63$^{+0.06}_{-0.05}$ & - & 26.42$^{+0.06}_{-0.05}$ \\
		6415 & GLASS & - & >29.88 & 28.78$^{+0.24}_{-0.19}$ & 28.63$^{+0.17}_{-0.14}$ & 28.43$^{+0.14}_{-0.12}$ & 29.09$^{+0.27}_{-0.22}$ & - & 28.85$^{+0.15}_{-0.13}$ \\
		17487 & GLASS & - & >29.69 & 29.52$^{+0.53}_{-0.35}$ & 26.83$^{+0.06}_{-0.05}$ & 27.17$^{+0.06}_{-0.05}$ & 27.30$^{+0.06}_{-0.06}$ & - & 27.31$^{+0.07}_{-0.07}$ \\
	\end{tabular}
	\label{tab:jwst_phot}
\end{table*}

\section{JWST Zero-point calibrations}

Since the initial release of the JWST imaging there have been updates
to the NIRCam calibrations from in-flight tests \citep{rigby2022}.
However, comparison to existing imaging data (both HST and ground based)
shows that there are still offsets in the NIRCam zero-points which differ
between each filter, module and sub-module.

In order to address this problem, we exploited existing imaging data covering the
CEERS survey field to derive filter and sub-module specific flux corrections that were
applied to our JWST photometric catalogue before the candidate selection process.

We explored two independent approaches to deriving the necessary flux corrections.
The first approach was based on SED fitting to a sample of objects with robust spectroscopic
redshifts from the DEEP2/3 spectroscopic survey \citep{cooper2012,newman2013}. By fixing the redshift to its spectroscopic value, the
difference between the observed fluxes in each filter and the fluxes predicted by the best-fitting SED template
provided one estimate of the average flux corrections in each filter/sub-module.

The second approach was based on directly comparing the observed NIRCam fluxes of compact objects
with fluxes measured from existing imaging data taken in overlapping filters (e.g. F125W, F160W, K$_{\rm s}$, IRAC CH1, IRAC CH2).
During this process, special attention was paid to adopting photometric apertures which enclosed the same fraction of total
flux and correcting for average colour terms between filters.

It is clear that different potential biases and systematic errors are likely to affect both
methods employed to derive the flux correction factors. Consequently, for the F115W, F150W, F200W, F277W,
F356W, F410M and F444W filters we adopted the straight average of the correction factors provided by the two different approaches.

The exception to this rule is the F090W filter, which does not form part of the CEERS data set. The adopted
correction factors for this filter were taken from a separate analyis \footnote{https://github.com/gbrammer/grizli/pull/107}. The flux correction factors
applied to our JWST photometric catalogue are listed in Table \ref{tab:ZP_offsets}. 

\begin{table}
	\centering
	\caption{Zero-point corrections applied to catalogue fluxes for each sub-module within each NIRCam filter. }
	\label{tab:ZP_offsets}
    \setlength{\tabcolsep}{4pt} 
	\renewcommand{\arraystretch}{1.15} 
	\begin{tabular}{lcc} 
		\hline
		Filter & Detector & Correction\\
		\hline
		F090W & NRCA1 & 0.83\\
		F090W & NRCA2 & 0.83\\
		F090W & NRCA3 & 0.76\\
		F090W & NRCA4 & 0.75\\
		F090W & NRCB1 & 0.89\\
		F090W & NRCB2 & 0.84\\
		F090W & NRCB3 & 0.91\\
		F090W & NRCB4 & 0.82\\
		\hline
		F115W & NRCA1 & 0.88\\
		F115W & NRCA2 & 0.88\\
		F115W & NRCA3 & 0.90\\
		F115W & NRCA4 & 0.86\\
		F115W & NRCB1 & 0.91\\
		F115W & NRCB2 & 0.88\\
		F115W & NRCB3 & 0.94\\
		F115W & NRCB4 & 0.85\\
		\hline
		F150W & NRCA1 & 0.90\\
		F150W & NRCA2 & 0.91\\
		F150W & NRCA3 & 0.92\\
		F150W & NRCA4 & 0.90\\
		F150W & NRCB1 & 0.93\\
		F150W & NRCB2 & 0.91\\
		F150W & NRCB3 & 0.95\\
		F150W & NRCB4 & 0.87\\
		\hline
		F200W & NRCA1 & 0.88\\
		F200W & NRCA2 & 0.88\\
		F200W & NRCA3 & 0.87\\
		F200W & NRCA4 & 0.88\\
		F200W & NRCB1 & 0.88\\
		F200W & NRCB2 & 0.88\\
		F200W & NRCB3 & 0.91\\
		F200W & NRCB4 & 0.88\\
		\hline
		F277W & NRCALONG & 1.05\\
		F277W & NRCBLONG & 0.97\\
		\hline
		F356W & NRCALONG & 1.06\\
		F356W & NRCBLONG & 1.01\\
		\hline
		F410M & NRCALONG & 0.99\\
		F410M & NRCBLONG & 1.01\\
		\hline
		F444W & NRCALONG & 1.07\\
		F444W & NRCBLONG & 1.05\\

		\hline
	\end{tabular}
\end{table}

\bsp
\label{lastpage}
\end{document}